\PassOptionsToPackage{table,xcdraw}{xcolor}
\documentclass[11pt,a4paper]{article}
\pdfoutput=1

\makeatletter
\def\@fpheader{}
\makeatother

\usepackage{jheppub}
\usepackage{gensymb}
\DeclareSymbolFont{matha}{OML}{txmi}{m}{it}% txfonts
\DeclareMathSymbol{\varv}{\mathord}{matha}{118}
\usepackage{amsmath}
\usepackage{amssymb}
\usepackage{graphicx}
\usepackage{subfigure}
\usepackage{amsfonts}
\usepackage{pstricks}
\usepackage{bm}
\usepackage{pbox}
\usepackage{placeins}
\usepackage{booktabs}
\usepackage[T1]{fontenc}
\usepackage{footnote}
\usepackage{pdfpages}
\usepackage{hhline}
\usepackage{multirow}
\usepackage{multicol}
\usepackage{enumitem}
\usepackage[toc,page]{appendix}
\allowdisplaybreaks
\usepackage{comment}
\usepackage{float}
\usepackage{slashed}
\usepackage{xcolor}
\usepackage{textcomp}
\usepackage{multirow}
\usepackage[numbers]{natbib}
\usepackage{notoccite}
\usepackage{indentfirst}
\usepackage{jheppub}
\usepackage{gensymb}
\DeclareSymbolFont{matha}{OML}{txmi}{m}{it}% txfonts
\DeclareMathSymbol{\varv}{\mathord}{matha}{118}
\usepackage{amsmath}
\usepackage{array}
\usepackage{amssymb}
\usepackage{graphicx}
\usepackage{subfigure}
\usepackage{pstricks}
\usepackage{bm}
\usepackage{pbox}
\usepackage{placeins}
\usepackage{booktabs}
\usepackage[T1]{fontenc}
\usepackage{footnote}
\usepackage{pdfpages}
\usepackage{hhline}
\usepackage{multirow}
\usepackage{multicol}
\usepackage{enumitem}
\usepackage[toc,page]{appendix}
\allowdisplaybreaks
\usepackage{comment}
\usepackage{float}
\usepackage{slashed}
\usepackage{xcolor}
\usepackage{textcomp}
\usepackage{multirow}
\usepackage[numbers]{natbib}
\usepackage{notoccite}
\usepackage{soul}
\usepackage{extarrows}

\usepackage{physics}
\usepackage{comment}
\usepackage{dsfont}
\usepackage{blindtext}
\usepackage{csquotes}
\usepackage{soul}
\usepackage[none]{hyphenat}
\usepackage{indentfirst}
\DeclareUnicodeCharacter{0304}{ }
\usepackage{rotating}
\usepackage{tabularx}
\usepackage{extarrows}
\usepackage{physics}
\usepackage{comment}
\usepackage{dsfont}
\usepackage{blindtext}
\usepackage{csquotes}
\usepackage[none]{hyphenat}
\DeclareUnicodeCharacter{0304}{ }
\usepackage{rotating}
\usepackage{tabularx}
\usepackage[many]{tcolorbox}
\definecolor{fg}{RGB}{34,139,34}

\renewcommand{\thetable}{\Roman{table}}

\def\figureautorefname~#1\null{Fig.\,#1\null}
\def\equationautorefname~#1\null{Eq.\,(#1)\null}
\def\tableautorefname~#1\null{Tab.\,#1\null}
\definecolor{MyDarkBlue}{rgb}{0.1, 0.1, 0.8} 
\definecolor{MyLightBlue}{rgb}{0.22,0.51,0.9}
\definecolor{MyGreen}{rgb}{0.0, 0.5, 0.0}
\definecolor{BrickRed}{rgb}{0.8, 0.25, 0.33}

\RequirePackage{hyperref}
\hypersetup{colorlinks=true, citecolor=MyGreen,linkcolor=BrickRed, urlcolor=MyLightBlue}
\title{\bf Dirac Leptogenesis in Left-Right Symmetric Models}
\author[a]{K.S. Babu,}
\author[a]{Ajay Kaladharan}
\affiliation[a]{Department of Physics, Oklahoma State University, Stillwater, OK 74078, USA}
\emailAdd{babu@okstate.edu}\emailAdd{kaladharan.ajay@okstate.edu}

\abstract{ Left-right symmetric models which employ a generalized seesaw mechanism to generate quark and charged lepton masses are known to solve the strong CP problem via parity symmetry, without the need for the axion.  These models lead to naturally light Dirac neutrinos with their masses arising through radiatve corrections. In this work, we show how baryogenesis via Dirac leptogenesis can be implemented in this framework.  A pair of left-right symmetric scalar doublets $\phi_{L,R}$ carrying $(B-L)$ charge of $3$ is introduced for this purpose, which preserves the parity solution to the strong CP problem.  Small neutrino masses are generated through one-loop diagrams mediated by these scalars.  The decay of vector-like leptons ($E_i$) involved in the seesaw mechanism to generate charged lepton masses, $E_i \rightarrow \overline{\nu}_{jR} \,\phi_R^-$, creates a right-handed neutrino ($\nu_{R}$) asymmetry, while preserving total lepton number.  The compensating lepton asymmetry is converted to baryon asymmetry via electroweak sphalerons.  We show that for a large range of model parameters successful baryogenesis can be realized, with the $W_R^\pm$ gauge boson mass of order $10^{11}$ GeV.
}

\begin{document}
\maketitle

\begin{sloppypar}

%%%%%%%%%%%%%%%%%%%%%%%%%%%%%%%%%%%%%%%%%%%%%%%%%%%%
\section{Introduction}
\label{sec:intro}

The Standard Model (SM) has been remarkably successful in explaining the fundamental forces of nature, but it falls short in explaining the matter-antimatter asymmetry~\cite{Planck:2018vyg} of the Universe as well as the experimentally observed neutrino oscillation data~\cite{Davis1994ARO,Super-Kamiokande:1998kpq,SNO:2002tuh,  DayaBay:2012fng} which require neutrinos to have small but non-vanishing masses. 
The oscillation data, however,  cannot differentiate
between the  Dirac nature  and the Majorana  nature of neutrinos.  If the neutrino is a Dirac particles, lepton number ($L$) would be a conserved symmetry, while for a Majorana neutrino $L$ is broken by two units. Observation of neutrinoless double beta decay, which violates $L$ by two units, would confirm the Majorana nature of neutrinos (for a review of the current status, see Ref.~\cite{Dolinski:2019nrj}). However, this process, or any other $L$-violating process has not been observed thus far, and the Dirac/Majorana nature of the neutrino remains an experimentally unsettled question.

In the case of Majorana neutrinos the seesaw mechanism~\cite{Minkowski:1977sc,Yanagida:1979as,Glashow:1979nm,Gell-Mann:1979vob,Mohapatra:1979ia} involving right-handed neutrinos $(\nu_R)$ explains the smallness of neutrino mass elegantly.  The same $\nu_R$ fields can be responsible for the baryon asymmetry of the Universe, as their decays into leptons and the Higgs doublet of the SM can generate an asymmetry in lepton number (``leptogenesis")~\cite{Fukugita:1986hr}, which is converted to baryon asymmetry by the non-perturbative sphaleron interactions of the electroweak sector~\cite{Kuzmin:1985mm}. (For a review see, for e.g, Ref.~\cite{Davidson:2008bu}.) Thus, two of the missing ingredients of the SM, viz., small but non-zero neutrino masses and a mechanism to create the baryon asymmetry of the Universe are solved with one stroke in this setup.

There is a similar, although less studied, mechanism for baryon asymmetry generation if the neutrinos are Dirac particles.  In this ``Dirac leptogenesis" mechanism~\cite{Dick:1999je}, an asymmetry in $\nu_R$ number is created at high temperatures in the early history of the Universe, without breaking lepton number.  This asymmetry survives down to low temperatures, since the $\nu_R$ fields remain out of equilibrium with the SM plasma.  The $\nu_R$'s, being singlets of the SM, interact with the plasma only through their tiny Yukawa couplings, which are too weak (given the smallness of the neutrino mass) to maintain equilibrium.  Since lepton number $L$ is conserved, an equal and opposite asymmetry is created in the SM sector, which is converted to baryon asymmetry by the electroweak sphaleron interactions. For realistic models that realize this mechanism in various extensions of the SM, see Ref.~\cite{Murayama:2002je,Boz:2004ga,Thomas:2005rs,Cerdeno:2006ha,Thomas:2006gr,Gu:2006dc, Gu:2007mc,Chun:2008pg,Bechinger:2009qk,Borah:2016zbd, Earl:2019wjw,Heeck:2023soj,Barman:2023fad}. Thus, this setup also solves the two shortcomings (small neutrino mass and baryon asymmetry) of the SM with a single stroke. One difference, however, compared to the Majorana leptogenesis is that the smallness of the neutrino mass is not always guaranteed in the case of Dirac neutrinos.  The purpose of this paper is propose and analyze Dirac leptogenesis in a class of left-right symmetric models where the neutrinos are naturally light Dirac particles with their small masses arising from radiative corrections.

Left-right symmetric models (LRSM) based on the gauge group $SU(3)_c \otimes SU(2)_L \otimes SU(2)_R \otimes U(1)_{B-L}$~\cite{Pati:1974yy, Mohapatra:1974hk, Mohapatra:1974gc, Senjanovic:1975rk, Mohapatra:1980yp} are well-motivated extensions of the SM, since they provide a dynamical understanding of the origin of parity violation.  These models require the existence of the right-handed neutrino $\nu_R$ as the $SU(2)_R$-partner of the right-handed electron ($e_R$) field, which makes them an excellent framework to address neutrino masses and their Majorana versus Dirac nature.  In a class of left-right symmetric models employing a simple Higgs sector consisting of a pair of $SU(2)_{L,R}$ doublets, the masses of quarks and charged leptons arise through a generalized seesaw mechanism involving vector-like fermions~\cite{Berezhiani:1983hm,Davidson:1987mh,Rajpoot:1987fca,Babu:1988mw,Babu:1989rb}. In this class of models it is natural for the neutrino to be a Dirac particle, as its mass is zero at the tree-level, with small masses arising through loop diagrams~\cite{Babu:1988yq,Babu:2022ikf,Mohapatra:1987nx}. The matter spectrum of the model fits in precisely in the simplest left-right symmetric $SU(5)_L \otimes SU(5)_R$ unified theory~\cite{Babu:2023dzz}.
The focus of this paper is the realization of  Dirac leptogenesis in this class of models.  A $\nu_R$ asymmetry is created at high temperatures in the early history of the Universe in the decay of the vector-like lepton $E_i$ involved in the charged lepton mass generation, which survives down to low temperatures, and is converted first to SM lepton asymmetry and then to baryon asymmetry by the electroweak sphalerons. 

This class of models has an added benefit in that it also provides a solution to the strong CP problem with parity symmetry, without the need for the axion~\cite{Babu:1989rb}.  The $\theta$ parameter of QCD is zero due to parity, and the structure of the quark mass matrix is such that the flavor contribution to $\overline{\theta}$ is also zero at tree-level as well as one-loop level.  The two-loop induced $\overline{\theta}$ is naturally small and consistent with neutron electric dipole moment constraints~\cite{Babu:1989rb,Banno:2023yrd,deVries:2021pzl}. These models have been extensively studied in the recent literature in the context of flavor physics~\cite{Craig:2020bnv,Dcruz:2022rjg,Dcruz:2023mvf,Jana:2021tlx},  neutrino oscillations~\cite{Borah:2017leo,Babu:2023dzz,Mohapatra1988LeftrightSA} and gravitational wave detection~\cite{Graf:2021xku}. The model we propose is an extension of the minimal model of Ref.~\cite{Babu:1989rb} with the addition of a pair of $SU(2)_{L,R}$ Higgs doublets carrying $(B-L)$ charge of 3, which serves two purposes. These scalars contribute to one-loop neutrino mass generation and also enable a mechanism to generate $\nu_R$ asymmetry in the early universe.

The rest of the paper is organized as follows. 
In the~\autoref{sec:model}, we introduce the details of the model and follow it with the neutrino mass mechanism in the~\autoref{sec:neutrino}. In~\autoref{sec:Lepto}, we explain the Dirac leptogenesis mechanism in the model and provide the relevant Boltzmann equations. In ~\autoref{sec:res}, we study the special case of resonant leptogenesis as applied to the model. Finally, in the~\autoref{sec:summ}, we summarize our main results. ~\autoref{sec:Yuk} provides the numerical values of the relevant Yukawa matrices and bare mass matrices for the benchmark points that we have presented.

%%%%%%%%%%%%%%%%%%%%%%%%%%%%%%%%%%%%%%%%%%%%%%%%%%%%%%%%%%%%
%%%%%%%%%%%%%%%%%%%%%%%%%%%%%%%%%%%%%%%%%%%%%%%%%%%%%%%%%%%%
\section{The Model}
\label{sec:model}
Left-right symmetric models are based on the gauge symmetry $SU(3)_c \otimes SU(2)_L \otimes SU(2)_R \otimes U(1)_{B-L}$~\cite{Pati:1974yy, Mohapatra:1974hk, Mohapatra:1974gc, Senjanovic:1975rk, Mohapatra:1980yp} . The standard model fermions, along with the right-handed neutrino, form doublets under this gauge group in a left-right symmetric manner ($i=1-3$ is the family index):
\begin{align}
    Q_{L,i}\left(3,2,1,+\frac{1}{3}\right) &\equiv \begin{pmatrix}u_L\\d_L\end{pmatrix}_i\,,&Q_{R,i} \left(3,1,2,+\frac{1}{3}\right)&\equiv\begin{pmatrix}u_R\\d_R\end{pmatrix}_i \,,
        % \label{eq:param_scan1}
    \nonumber \\
    \psi_{L,i} \left(1,2,1,-1\right)&\equiv\begin{pmatrix}\nu_L\\e_L\end{pmatrix}_i,\,& \psi_{R,i}\left(1,1,2,-1\right)&\equiv\begin{pmatrix}\nu_R\\e_R\end{pmatrix}_i.
     \label{eq:fermion_def}
\end{align}
Three families of vector-like fermions, which are singlets under $SU(2)_L \otimes SU(2)_R$, are introduced in the model to generate masses for the usual fermion via a generalized seesaw mechanism ~\cite{Berezhiani:1983hm,Davidson:1987mh,Rajpoot:1987fca,Babu:1988mw,Babu:1989rb}:
\begin{equation}
E_{L/R}(1,1,1,-2), \quad \quad U_{L/R}(3,1,1,\frac 43), \quad \quad D_{L/R}(3,1,1,-\frac 23).
\label{eq:vec-fermion}
\end{equation}

The specific version of this class of left-right symmetric model that we propose for small Dirac neutrino masses as well as Dirac leptogenesis has the following Higgs fields:
\begin{align}
    \chi_{L}\left(1,2,1,1\right) &\equiv \begin{pmatrix}\chi_L^+\\\chi_L^0\end{pmatrix}\,,&\chi_{R}\left(1,1,2,1\right) &\equiv\begin{pmatrix}\chi_R^+\\\chi_R^0\end{pmatrix} \,,
    \nonumber \\
    \phi_{L} \left(1,2,1,3\right)&\equiv\begin{pmatrix}\phi_L^{++}\\\phi_L^{+}\end{pmatrix},\,& \phi_{R}\left(1,1,2,3\right)&\equiv\begin{pmatrix}\phi_R^{++}\\\phi_R^{+}\end{pmatrix}.
     \label{eq:scalar_def}
\end{align}
The fields $\chi_{L,R}$ are employed to break the gauge symmetry down to ${SU}(3)_c\otimes {U}(1)_{\mathrm{em}}$. This spontaneous symmetry breaking occurs in two steps: $SU(3)_c\otimes SU(2)_L\otimes SU(2)_R\otimes U(1)_{B-L} \rightarrow SU(3)_C\otimes SU(2)_L\otimes U(1)_{Y} \rightarrow SU(3)_c\otimes U(1)_{\mathrm{em}}$, via vacuum expectation values (VEVs) of the neutral components of $\chi_R$ and $\chi_L$ given by
\begin{equation}
\left \langle \chi_L^0 \right \rangle =\kappa_L, \quad \quad 
\left \langle \chi_R^0 \right \rangle =\kappa_R,
\end{equation}
with the hierarchy $\kappa_R \gg \kappa_L = 174$ GeV. The $(B-L)= 3$ scalar fields $\phi_{L,R}$ are used to generate small Dirac neutrino masses via one-loop Feynman diagrams.  They also enable the creation of $\nu_R$ asymmetry in the early universe via the decay $E_i \rightarrow \overline{\nu}_{Rj} \phi_R^-$.  It should be noted that small Dirac neutrino masses can be induced in the model via the exchange of a singlet charged scalar (rather than the pair of doublets $\phi_{L,R}$ of~\autoref{eq:scalar_def}~\cite{Mohapatra1988LeftrightSA}).  But in such a scenario, we found that it is difficult to explain baryon asymmetry via Dirac leptogenesis and neutrino oscillation data simultaneously. In Ref.~\cite{Borah:2017leo}, a second pair of doublet scalars $\chi'_{L,R}$ fields with a $(B-L)$ charge of $1$ was proposed to generate Dirac neutrino masses through the exchange of the physical charged Higgs bosons.  In that scenario, however, the fact that both doublets have neutral components acquiring VEVs, which are complex in general, would spoil the strong CP solution via parity symmetry (the determinant of the quark mass matrix will no longer be real). Our choice of $\phi_{L,R}$ doublets as shown in~\autoref{eq:scalar_def} avoids this problem, as there are no neutral components in these fields that acquire VEVs.  Furthermore, these $\phi_{L,R}$ fields do not couple to the quarks at all, and the parity solution to the strong CP problem is preserved as in the minimal model~\cite{Babu:1989rb}. The exchange of $\phi_{R,L}^-$ fields generate small Dirac neutrino masses via one-loop diagrams in our model.\footnote{Ref.~\cite{Maharathy:2022gki} has introduced both a singlet charged scalar and the $(B-L)=3$ doublet scalars $\phi_{LR}$ to generate neutrino masses.  However, we find that the singlet charged scalar is not necessary for this purpose, as shown in~\autoref{sec:neutrino}.} The couplings of $\phi_{L,R}$ fields to the leptons is such that lepton number remains conserved, which can be seen by assigning $L = 2$ to $\phi_{L,R}$. (This will become clear from the Yukawa interactions discussed later in this section.) Owing to this lepton number symmetry $\phi_{L,R}^\pm$ do not mix with the $\chi_{L,R}^\pm$ fileds which carry $L = 0$ and are in fact the Goldstone bosons eaten by the $W_{L,R}^\pm$ gauge bosons.

The general Higgs potential involving the $\chi_{L,R}$ and $\phi_{L,R}$ fields is given by
\begin{align}
V=&-\mu_L^2\chi_L^\dagger\chi_L+\mu_R^2\chi_R^\dagger\chi_R+m_L^2\phi_L^\dagger\phi_L+m_R^2\phi_R^\dagger\phi_R
+\frac{\lambda_{1}}{2} \left ( (\chi_L^\dagger\chi_L)^2+(\chi_R^\dagger\chi_R)^2 \right ) \nonumber\\
&+\lambda_2(\chi_L^\dagger\chi_L)(\chi_R^\dagger\chi_R)+\frac{\lambda_{3}}{2}\left ((\phi_L^\dagger\phi_L)^2+(\phi_R^\dagger\phi_R)^2 \right )+\lambda_4(\phi_L^\dagger\phi_L)(\phi_R^\dagger\phi_R) \nonumber\\
&+\left \{ \lambda_5(\chi_L^\dagger\phi_L)(\phi_R^\dagger\chi_R) +\mathrm{h.c} \right \}+\lambda_{6} \left ((\chi_L^\dagger \chi_L)(\phi_L^\dagger \phi_L)+(\chi_R^\dagger \chi_R)(\phi_R^\dagger \phi_R) \right ) \nonumber\\
&+\lambda_{7}\left ((\chi_L^\dagger \phi_L)(\phi_L^\dagger \chi_L)+(\chi_R^\dagger \phi_R)(\phi_R^\dagger \chi_R)\right )+\lambda_{8} \left ( (\chi_L^\dagger \chi_L)(\phi_R^\dagger \phi_R)+(\chi_R^\dagger \chi_R)(\phi_L^\dagger \phi_L) \right ).
\end{align}
Here we have imposed parity symmetry under which $\chi_L \leftrightarrow \chi_R$ and $\phi_L \leftrightarrow \phi_R$, but we have allowed for its soft breaking by the quadratic mass terms for these fields.  

The $2 \times 2$ mass matrix for the neutral scalar fields $\sigma_{L/R}=\mathrm{Re}(\chi_{L/R}^0)$ is given by
\begin{equation} \mathcal{M}_{\sigma_{L,R}}=\begin{bmatrix}
2\lambda_{1}\kappa_L^2 &2\lambda_2 \kappa_L \kappa_R \\ 
 2\lambda_2 \kappa_L \kappa_R&2 \lambda_{1}\kappa_R^2
\end{bmatrix}.
\end{equation}
The lightest mass eigenstate (of mass $M_h$) is identified as the Standard Model Higgs $h$ of mass 125 GeV, while $H$ denotes a heavier scalar. The mass eigenstates and the physical Higgs masses are given by ($s_x\equiv \sin x,\, c_x\equiv \cos x$)
\begin{align}
h&=c_\xi \sigma_L-s_\xi \sigma_R,\quad\quad
H=s_\xi \sigma_L+c_\xi  \sigma_R\\
M^2_h&\simeq 2\lambda_{1}\left ( 1-\frac {\lambda_2^2}{\lambda_{1}^2} \right )\kappa_L^2,\quad \quad
M^2_H\simeq 2\lambda_{1}\kappa_R^2.
\end{align}
The mixing angle $\xi$ is given by
\begin{equation}
    \tan {2\xi}=\frac {2\lambda_2 \kappa_L\kappa_R}{\lambda_{1}\kappa_R^2-\lambda_{1}\kappa_L^2}.
\end{equation}

The masses of doubly charged scalars are given by
\begin{align}
M^2_{\phi_L^{++}}&=m^2_L+\lambda_{6}\kappa_L^2+\lambda_{8}\kappa_R^2, \\
M^2_{\phi_R^{++}}&=m^2_R+\lambda_{6}\kappa_R^2+\lambda_{8}\kappa_L^2.
\end{align}
The singly charged scalars $\phi_{L/R}^+$ mix with a mass matrix given by
\begin{equation}
   M^2_{\phi_{L/R}^+}=\begin{bmatrix}
m^2_L+(\lambda_{6}+\lambda_{7})\kappa_L^2+\lambda_{8}\kappa_R^2 & \lambda_5\kappa_L\kappa_R\\ 
\lambda_5\kappa_L\kappa_R & m^2_R+(\lambda_{6}+\lambda_{7})\kappa_R^2+\lambda_{8}\kappa_L^2
\label{eq:chargemass}
\end{bmatrix}.
\end{equation}
The mass eigenstates and the physical masses of these charged scalars are given by
\begin{align}
h_{1}^\pm&=c_{\theta_\pm} \phi_L^\pm-s_{\theta_\pm}\phi_R^\pm\\
h_{2}^\pm&=c_{\theta_\pm}\phi_R^\pm+s_{\theta_\pm}\phi_R^\pm\\
M^2_{h_{1/2}^\pm}&=\frac 12\left ( \mathcal{M}_{11}+\mathcal{M}_{11}\mp\sqrt{(\mathcal{M}_{22}-\mathcal{M}_{11})^2+4\mathcal{M}_{12}^2} \right ),
\end{align}
where $\mathcal{M}_{ij}$ denotes $(i,j)$ element of~\autoref{eq:chargemass} and the mixing angle $\theta_{\pm}$ is given by
\begin{equation}
\tan{2\theta_\pm}=\frac {2\mathcal{M}_{12}}{\mathcal{M}_{22}-\mathcal{M}_{11}},
\end{equation}
It is worth noting that in the electroweak symmetric vacuum, which is realized at temperatures well above 100 GeV, the mixing angle $\theta_\pm$ vanishes and therefore the mass eigenvalues are given by the diagonal entries of~\autoref{eq:chargemass}, which we shall denote as $M^2_{\phi_L^+}$ and 
 $M^2_{\phi_R^+}$, respectively.

The Yukawa couplings of fermions, along with the bare mass terms for the vector-like fermions, are given by the Lagrangian
\begin{align}
\mathcal{L}_{\mathrm{Yuk}}&=\mathcal{Y}_u\bar{Q}_L\tilde{\chi}_L U_R+\mathcal{Y}_u\bar{Q}_R\tilde{\chi}_R U_L+M_U\bar{U}_LU_R \nonumber\\
&+\mathcal{Y}_d\bar{Q}_L\chi_L D_R+\mathcal{Y}_d\bar{Q}_R\chi_R D_L+M_D\bar{D}_LD_R \nonumber\\
&+\mathcal{Y}_\ell\bar{\psi}_L\chi_L E_R+\mathcal{Y}_\ell\bar{\psi}_R\chi_R E_L+M_E\bar{E}_LE_R \nonumber\\
&+f\left ({\psi}_{Li}^T \mathcal{C}\phi_{Lj}E_L\epsilon^{ij}+{\psi}_{Ri}^T \mathcal{C}\phi_{Rj}E_R\epsilon^{ij}  \right )  +\mathrm{h.c},
\label{eq:Yuk}
\end{align}
with $\tilde{\chi}_{L/R}=i\tau_2\chi^*_{L/R}$ and $i,j$ are $SU(2)_{L/R}$ indices. Under parity symmetry the fermions and scalar fields transform as
\begin{equation}
Q_L \leftrightarrow Q_R, ~\psi_L \leftrightarrow \psi_R,~ U_L \leftrightarrow U_R, ~D_L \leftrightarrow D_R, ~E_L \leftrightarrow E_R, ~\chi_L \leftrightarrow \chi_R, ~\phi_L \leftrightarrow \phi_R~.
\end{equation}
As a result, the vector-like fermion mass matrices $M_{U,D,E}$ are hermitian in~\autoref{eq:Yuk}. The Lagrangian of \autoref{eq:Yuk} generates $6 \times 6$ mass matrices for charged leptons ($e, \, E$), up-type quarks ($u, \,\, U$), and down-type quarks ($d, \, D$), which can be expressed in block form as 
\begin{equation}
    \mathcal{M}_f=\begin{pmatrix}
0 &\mathcal{Y}_f \kappa_L\\ 
\mathcal{Y}^\dagger_f \kappa_R & M_F 
\end{pmatrix}~~~(f=\ell,\,u,\,d),\, \rm{and}\, (F=E,U,D).
\label{eq:Massma}
\end{equation}
It is clear from here that the determinant of the quark mass matrices are real, which helps solve the strong CP problem.

Using two unitary matrices $\mathcal{U}_{L/R}$, we can block-diagonalize matrix $\mathcal{M}_f$ via a bi-unitary transformation, $\mathcal{U}_L^\dagger \mathcal{M}_f \mathcal{U}_R = \mathcal{M}_{\rm block}$. Defining
\begin{equation}
\rho_{Lf}=\mathcal{Y}_f \kappa_L M_f^{-1}, \quad \quad \rho_{Rf}=\mathcal{Y}_f \kappa_R M_f^{-1},
\end{equation} 
we can parametrize the unitary matrices $\mathcal{U}_{L/R}$, with  the assumption $\rho_{L_f,R_f}\ll1$ as
\begin{equation}
\mathcal{U}_{Lf}=\begin{pmatrix}
1-\frac {1}2\rho_{Lf}^\dagger \rho_{Lf} & \rho_{Lf} \\ 
 -\rho_{Lf}^\dagger&  1-\frac {1}2\rho_{Lf} \rho_{Lf}^\dagger
\end{pmatrix}, \quad \quad
\mathcal{U}_{Rf}=\begin{pmatrix}
1-\frac {1}2\rho_{Rf}^\dagger \rho_{Rf} & \rho_{Rf} \\ 
 -\rho_{Rf}^\dagger&  1-\frac {1}2\rho_{Rf} \rho_{Rf}^\dagger
\end{pmatrix}.
\end{equation}
{  In this work, we focus on the leptonic sector and, for simplicity, we drop the explicit leptonic indices, henceforth referring to $\rho_{L_\ell, R_\ell}$ as $\rho_{L, R}$.} The assumption $\rho_L \ll 1$ follows from our phenomenological requirement $\kappa_L \ll \kappa_R$, so that the $W_R^\pm$ mass is much larger than the $W_L^\pm$ mass.  While $\rho_R$ can be of order unity in general, to simplify our analysis, and to realize Dirac leptogenesis, we will focus on the regime with $\rho_{L,R}\ll1$. 
This is because we are interested in the parameter space with $M_E$ is of similar order as $\kappa_R$.  To explain the charged lepton mass, one requires $\mathcal{Y}_\ell\ll 1$ in this case, which leads to $\rho_R \ll 1$. (For a $2 \times 2$ seesaw matrix of the form given in~\autoref{eq:Massma}, the lightest eigenvalue is $m_\ell \simeq y_\ell \kappa_L \cos\theta_R^\ell$, where $\tan\theta_R^\ell = y_\ell \kappa_R/M_\ell$.  When $\kappa_R \sim M_\ell$, $\cos\theta_R \sim \mathcal{O}(1)$ and to explain the charged lepton mass one would need $y_\ell \ll 1$, and hence $\rho_R \ll 1$.) We shall assume this in our numerical analysis, but we have verified that this condition holds by checking the approximate results with exact numerical results.

The $3 \times 3$ blocks of the block-diagonal matrix $M_{\rm block}$ are given by
\begin{equation}
\hat{m_f}=-\kappa_L\kappa_R\mathcal{Y}_f M_f^{-1}\mathcal{Y}_f^\dagger, \quad \quad 
\hat{M_f}=M_f+\frac 12\left ( \kappa_R^2\mathcal{Y}_f^\dagger\mathcal{Y}_f M_f^{-1} +\kappa_L^2\mathcal{Y}_f\mathcal{Y}_f^\dagger M_f^{-1} \right ).
\label{eq:masseig}
\end{equation}
The hermitian matrix $\hat{m}_f$ can be diagonalized using a unitary matrix $m_f=U_{f} \hat{m_f} U_{f}^\dagger$, while the matrix $\hat{M}$  can be diagonalized using a bi-unitary transformation $\hat{M}_f=V_{Lf} M_f^d V_{Rf}^\dagger$, where $m_f$ and $M_f^d$ are the diagonal matrix. In the case when $M_f$ is diagonal, which can be chosen without loss of generality by a flavor rotation, the Yukawa matrix $\mathcal{Y}_f$ can be parametrized as
\begin{equation}
\mathcal{Y}_{f} = \frac {1}{\sqrt{\kappa_L \kappa_R}}U_f^\dagger m_f^{1/2} U_f'  M_f^{1/2},
\end{equation}
where $U_f'$ is general unitary matrix which can be parameterized using the standard CKM-like parameterization.

The charged gauge bosons do not undergo mixing at the tree level, and their masses are evaluated as:
\begin{equation}
{M^2}_{W^\pm_{L}}=\frac 12 g_{L}\kappa_{L}^2, \quad \quad {M^2}_{W^\pm_{R}}=\frac 12 g_{R}\kappa_{R}^2.
\end{equation}
For the neutral gauge boson sector, the photon field $A_\mu$ remains massless while the orthogonal fields $Z_L$ and $Z_R$ undergo mixing at the tree level, and the mixing matrix is given by
\begin{equation}
M^2_{Z_L-Z_R}=\frac 12 
\begin{pmatrix}
 (g_L^2+g_Y^2)\kappa_L^2 & g_Y^2\sqrt{\frac {g_L^2+g_y^2}{g_R^2-g_Y^2}}\kappa_L^2\\ 
 g_Y^2\sqrt{\frac {g_L^2+g_y^2}{g_R^2-g_Y^2}}\kappa_L^2 & \frac {g_R^4\kappa_R^2+g_Y^2\kappa_L^2}{g_R^2-g_y^2}
\end{pmatrix}.
\end{equation}
The masses of neutral gauge bosons are given approximately by
\begin{equation}
M^2_{Z_1}\simeq \frac 12 (g_Y^2+g_L^2)\kappa_L^2\, \quad \quad \quad
M^2_{Z_2}\simeq \frac 12 \frac {g_R^4\kappa_R^2+g_Y^2\kappa_L^2}{g_R^2-g_y^2}.
\end{equation}
Under the parity symmetry, the gauge boson transforms as $W_{L}^\pm \leftrightarrow W_R^\pm$, implying that the gauge couplings for $SU(2)_L$ and $SU(2)_R$ are equal: $g_L=g_R$ at the scale $\kappa_R$.
The embedding of $U(1)_Y\subset SU(2)_R \otimes U(1)_{B-L}$ yields the matching condition for hypercharge $Y$ and the corresponding gauge coupling $g_Y$,
\begin{equation}
\frac {Y}{2}=T_{3R}+\frac {B-L}{2}, \quad\quad\quad \frac {1}{g_Y^2}=\frac {1}{g_R^2}+\frac {1}{g_{B-L}^2}.
\end{equation}
These relations will be used in our numerical studies of Dirac leptogenesis.

It is important to emphasize that the neutrino masses are zero at the tree level in the model.  This is partly because there is no vector-like neutral lepton with which the neutrino can mix. In fact, the vector-like fermion spectrum given in~\autoref{eq:vec-fermion} is exactly what would arise from an embedding of the left-right theory in a parity-symmetric $SU(5)_L \otimes SU(5)_R$ unified theory, which justifies the absence of vector-like neutral leptons.  The absence of tree-level neutrino mass also has to do with the absence of a bidoublet scalar used in the standard LRSM, which is not necessary in the present context of generalized seesaw for quark and charged lepton mass generation.  Neutrinos will acquire small Dirac masses through loop diagrams, which we analyze in the next section.

%%%%%%%%%%%%%%%%%%%%%%%%%%%%%%%%%%%%%%%%%%%%%%%%%%%%%%%%%%%%
\section {Loop-Induced Dirac Neutrino Masses}
\label{sec:neutrino}

In the model under study the neutrinos can be naturally light Dirac particles.  As argued earlier, there is no vector-like neutral lepton that can induce the neutrino a tree-level mass.  While it is possible that without additional scalar fields the neutrino would acquire small Dirac mass via a two-loop induced diagram mediated by $W_L^+-W_R^+$ mixing~\cite{Babu:1988yq,Babu:2022ikf}, consistent with neutrino oscillation data, that scenario does not apear to have the ingredients for successful Dirac leptogenesis.   Our proposal extends the simplest model with the pair of $B-L=3$ scalar fields $\phi_{L,R}$, which we found to be the simplest extension to accommodate  Dirac leptogenesis, as well as to explain neutrino oscillation data. In this context, we shall assume that the contribution of the two-loop $W_L^{\pm}-W_R^{\pm}$ exchange diagram is small, which can be achieved, for example, by keeping the bare mass of the top partner $M_{U_3}$ to be small, so that the loop-induced $W_L^{\pm}-W_R^{\pm}$ mixing is suppressed.

To evaluate the loop-induced Dirac neutrino masses, we work in a basis where the vector-like charged lepton mass matrix $M_E$ is diagonal.  
\begin{figure}[th!]
\centering
\includegraphics[width=0.8\textwidth]{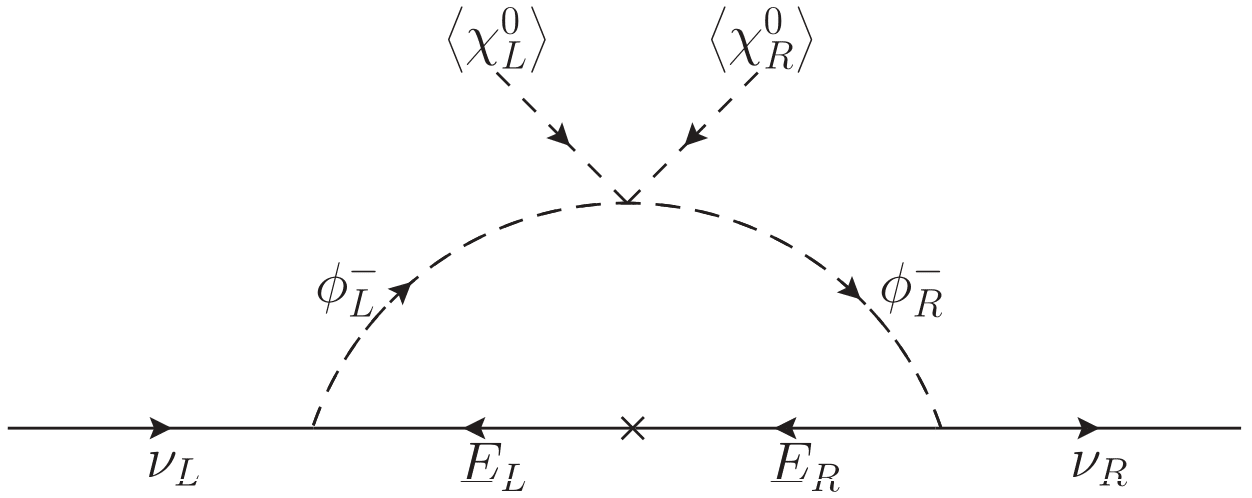}
\caption{Diagram generating tiny Dirac neutrino masses. }
\label{fig:Nmass}
\end{figure}
The model conserves lepton number; the neutrinos acquire tiny Dirac masses via a one-loop mechanism arising from the diagram in the~\autoref{fig:Nmass}. 
The induced mass matrix can be calculated in analogy to a similar Dirac neutrino mass model analyzed in Ref.~\cite{PhysRevD.86.033007}, following a procedure akin to that of the one-loop Majorana mass~\cite{ZEE1980389,Ma:2006km}. The quartic coupling $\lambda_5$  induces mixing between $\phi_{L}^\pm$ and $\phi_{R}^\pm$, with $\lambda_5$ taken  to be real without loss of generality.  The neutrino masses arise when both $\chi_L^0$ and $\chi_R^0$ acquire non-zero vacuum expectation values. In the $\rho_{R,L}\ll 1$ limit, the neutrino mass matrix can be approximated as
\begin{equation}
({m_{\nu}})_{\alpha \beta}\simeq \sum_i\left ( \frac {(f_L)_{\alpha i}(f_R)^\star_{\beta i }s_{2\theta_\pm}}{32\pi^2}\left [ \frac {M_{E_i}M^2_{h_2^\pm}}{M^2_{h_2^\pm}-M^2_{E_i}}\ln\left ( \frac {M^2_{h_2^\pm}}{M^2_{E_i}} \right )-\frac {M_{E_i}M^2_{h_1^\pm}}{M^2_{h_1^\pm}-M^2_{E_i}}\ln\left ( \frac {M^2_{h_1^\pm}}{M^2_{E_i}} \right ) \right ] \right ),
\end{equation}
where $f_R=f V_{RE}$ and $f_L=f V_{LE}$.  Here, $V_{LE}$ and $V_{RE}$ are the unitariy transformations that diagonalizes the bare mass matrix of the heavy leptons, such that $\hat{M}_\ell=V_{LE} M_\ell^d V_{RE}^\dagger$. In the $\rho_L, \, \rho_R \ll 1$ limit, the matrix $\hat{M}_\ell$ is approximately hermitian, hence $V_{LE} \simeq V_{RE}$.  In a basis where the charged lepton mass matrix is diagonal, the Dirac neutrino mass matrix takes the form
\begin{equation}
\tilde{m}_\nu=U_{ \ell} m_{\nu} U_{ \ell}^\dagger.
\label{eq:mtilde}
\end{equation}
Subsequently, the neutrino mass matrix can be diagonalized by a unitarity transformation:
\begin{equation}
m_\nu^d=U^\dagger \tilde{m}_\nu U,
\end{equation}
where $U$ is the usual PMNS matrix.
For a given value of $M_{h_{1,2}^\pm}$ and $M_{E_i}$, the Dirac neutrino masses and mixing depend on the Yukawa matrix $f$ and the mixing angle $\theta_\pm$. This allows enough freedom in the parameter space to provide excellent fits for the neutrino oscillation data. The smallness of the Dirac mass can be explained, for example, by choosing $f \sim \lambda_5 \sim 10^{-3}$, $\kappa_R/M_E \sim 0.1$, which would leads to an effective Dirac Yukawa coupling of order $10^{-12}$. 
The mixing angle $\theta_\pm$ and Yukawa matrix $f$ play a crucial role in the Dirac leptogenesis mechanism, as we discuss in the~\autoref{sec:Lepto} later, introducing additional constraints on the parameter space. We shall present self-consistent fits to the neutrino oscillation data by also including these Dirac leptogenesis constraints in~\autoref{sec:Lepto}.

\section{Dirac Leptogenesis}
\label{sec:Lepto}
\begin{figure}[th!]
\centering
\includegraphics[width=\textwidth]{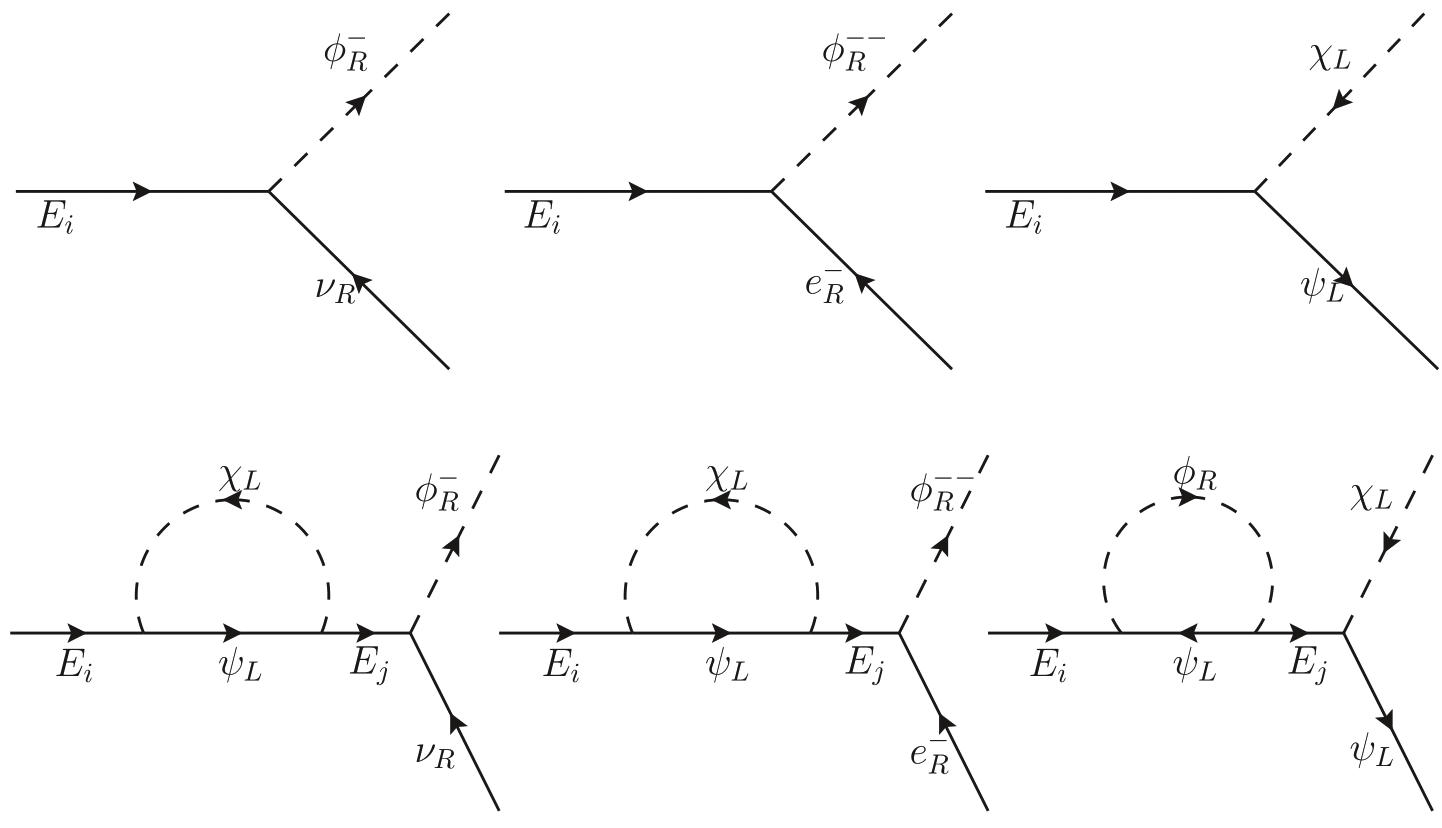}
\caption{The tree level and one-loop wave correction for the decay of heavy Dirac state $E_i$ to $\bar{\nu}_R\phi_R^-$, $e_R^+\phi_R^{--}$ and $\psi_L \bar{\chi}_L$. This diagram results in non-zero CP violation and subsequently contributes to baryon asymmetry.}
\label{fig:assym}
\end{figure}
In the model, Dirac leptogenesis occurs after the breaking of $SU(2)_R\otimes U(1)_{B-L}$.  This leptogenesis mechanism requires a mass spectrum where $M_{E_1}>M_{\phi_R^\pm}$. The left- and right-handed massive charged leptons combine to form heavy Dirac states $E\equiv {E}_R+{E}_L$. In analogy to standard leptogenesis, the decay of the heavy Dirac state, as shown in~\autoref{fig:assym}, could induce the CP violation and departure from thermal equilibrium. This scenario differs from other Dirac leptogenesis models~\cite{Cerdeno:2006ha, Murayama:2002je, Dick:1999je} as it occurs in two steps, described below.

In the first step, the asymmetry in $\nu_R$ is generated through the decay  of the heavy Dirac state $E_i$:
\begin{equation}
E_i\rightarrow \bar{\nu}_R\phi_R^-.
\end{equation}
The charged scalar $\phi_R^-$ carries lepton number $L = 2$. In the process of generating asymmetry in $\nu_R$, an equal and opposite asymmetry is simultaneously produced in $\phi_R^-$. To transfer this asymmetry into Standard Model leptons, $\phi_R^-$ must decay into $\phi_L^-$, which subsequently decays into SM leptons. 
\begin{figure}[th!]
\centering
\includegraphics[width=\textwidth]{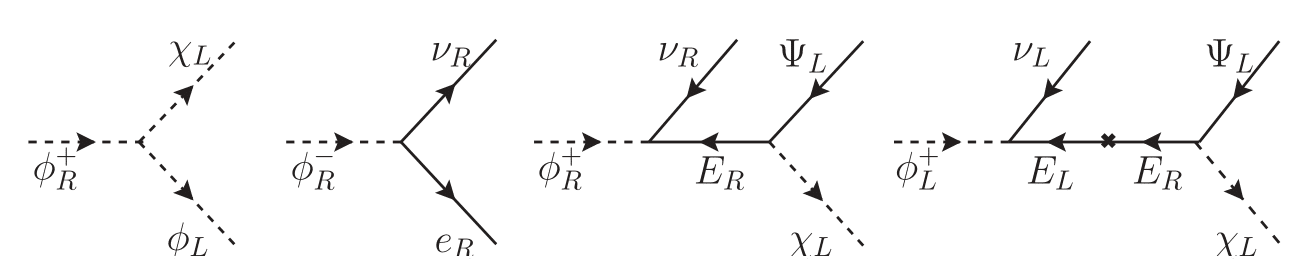}
\caption{The decay channel for the charged scalars $\phi_L^+$ and $\phi_R^+$.}
\label{fig:decay}
\end{figure}
{ The second step involves the decay of $\phi_R^-$, which proceeds through three channels shown in~\autoref{fig:decay}: the two body decays $\phi_R^- \rightarrow \phi_L \bar{\chi}_L$, $\phi_R^- \rightarrow \nu_R e_R^-$ and the three-body decay $\phi_R^- \rightarrow \nu_R \psi_L \bar{\chi}_L$. }To transfer the asymmetry into Standard Model leptons, the decay $\phi_R^- \rightarrow \phi_L \bar{\chi}_L$ must remain in equilibrium, while the other two channels should be out-of-equilibrium at $T \sim M_{\phi_R^-}$. Additionally, $\phi_L$ must decay into SM fermions before electroweak symmetry breaking. Due to the smallness of the Dirac neutrino mass, the asymmetry in $\nu_R$ is preserved, which in turn ensures the preservation of asymmetry in the Standard Model leptons, consistent with lepton number conservation.

The total decay width of the Dirac states $E_i$ can be defined by\footnote{{ We have assumed the mass of heavy Higgs $M_H \approx \kappa_R$, which kinematically forbids the decay $E_i \to \psi_R \bar\chi_R^0$.}}
\begin{align}
\Gamma(E_i)\equiv\Gamma_i=&\sum_\alpha \left (\Gamma(E_i\rightarrow\bar{\nu}_{R\alpha}\phi_R^-)+\Gamma(E_i\rightarrow\bar{e}_{R\alpha}\phi_R^{--})+\Gamma(E_i\rightarrow\bar{\psi}_{L\alpha}\phi_L)+\Gamma(E_i\rightarrow{\psi}_{L\alpha}\bar{\chi}_L)  \right ),\\
\Gamma_i=&\frac{{M}_{E_i}}{16\pi}\left ( ( f_L^\dagger f_L)_{ii}+(f_R^\dagger f_R)_{ii}+( \mathcal{Y}_L^\dagger\mathcal{Y}_L )_{ii} \right ),
\label{eq:gammai}
\end{align}
where $\mathcal{Y}_L=\mathcal{Y}_\ell V_{LE}$ and we assumed scalar particle masses are much less that $M_{E_i}$, neglecting the scalar mass dependence in \autoref{eq:gammai}. The tree-level branching ratios of the heavy lepton decaying into various channels are given by:
\begin{align}
\mathrm{Br}^i_{\bar{\nu}_R\phi_R^-}&=\frac {\Gamma(E_i\rightarrow\bar{\nu}_{R\alpha}\phi_R^-)}{\Gamma_i}=\frac {(f_R^\dagger f_R )_{ii}}{2(f_R^\dagger f_R )_{ii}+2( f_L^\dagger f_L )_{ii}+2(\mathcal{Y}_{ L}^\dagger\mathcal{Y}_L )_{ii}}\\
 \mathrm{Br}^i_{\bar{e}_R\phi_R^{--}}&=\frac {\Gamma(E_i\rightarrow\bar{e}_{R\alpha}\phi_R^{--})}{\Gamma_i}=\frac {(f_R^\dagger f_R )_{ii}}{2(f_R^\dagger f_R )_{ii}+2( f_L^\dagger f_L )_{ii}+2(\mathcal{Y}_{ L}^\dagger\mathcal{Y}_L )_{ii}}\\
  \mathrm{Br}^i_{\bar{\psi}_L\phi_L}&=\frac {\Gamma(E_i\rightarrow\bar{\psi}_L\phi_L)}{\Gamma_i}=\frac {2(f_L^\dagger f_L )_{ii}}{2(f_R^\dagger f_R )_{ii}+2( f_L^\dagger f_L )_{ii}+2(\mathcal{Y}_{ L}^\dagger\mathcal{Y}_L )_{ii}}\\
  \mathrm{Br}^i_{\psi_L\bar{\chi}_L}&=\frac {\Gamma(E_i\rightarrow\psi_L\bar{\chi}_L)}{\Gamma_i}=\frac {2(\mathcal{Y}_{ L}^\dagger\mathcal{Y}_L )_{ii}}{2(f_R^\dagger f_R )_{ii}+2( f_L^\dagger f_L )_{ii}+2(\mathcal{Y}_{ L}^\dagger\mathcal{Y}_L )_{ii}}
\label{eq:Bri}
\end{align}

The CP asymmetry associated with the process $E_i \rightarrow XY$ is defined as,
\begin{equation}
\epsilon_{XY}^i=\frac {\Gamma(E_i^-\rightarrow XY)-\Gamma(\bar E_i^+\rightarrow \bar{X}\bar{Y})}{2\Gamma_i}.
\end{equation}
Due to the CPT invariance, the decay width of Dirac state $E_i$ and its conjugate are equal: $\Gamma_i=\bar{\Gamma}_i$. As a consequence, the sum of CP asymmetry due to the decay of $E_i$ vanishes:
\begin{equation}
\epsilon^i_{\bar{\nu}_R \phi_R^-}+\epsilon^i_{e_R^+ \phi_R^{--}}+\epsilon^i_{\bar{\psi}_L \phi_L}+\epsilon^i_{\psi_L\bar{\chi}_L }=0.
\end{equation}

The diagrams responsible for the CP violation are provided in~\autoref{fig:assym}. There is no CP violation in the $\bar{\psi_L}\phi_L$ decay channel. The one-loop diagram with amplitude proportional to $f^2$ and $\mathcal{Y}_\ell^2$ does not lead to CP asymmetry, as summing over flavor indices cancels out CP violation in these diagrams.{\footnote{If the propagator of the internal $E_j$ picks up the momentum term instead of the mass term in the evaluation of CP asymmetry from the interference between tree-level and one-loop diagrams, the CP asymmetry vanishes. Moreover, it is crucial to emphasize that if the same Yukawa interaction appears in both the tree-level and one-loop diagrams, then summing over flavor indices results in a CP asymmetry proportional to either $\mathrm{Im}\left [ (\mathcal{Y}_L^\dagger \mathcal{Y}_L)_{ki}(\mathcal{Y}_R^\dagger \mathcal{Y}_R)_{ik} \right ]$ or $\mathrm{Im}\left [ (f_L^\dagger f_L)_{ki}(f_R^\dagger f_R)_{ik} \right ]$,
which vanishes in the limit $V_L \approx V_R$. The combination of these two factors leads to the absence of CP asymmetry in the $\bar{\psi}_L \phi_L$ decay channel.}}
The CP asymmetries generated by the decay of heavy Dirac state $E_i$ is given by:
\begin{align}
\epsilon^i_{\bar{\nu_R}\phi_R^-}&=\frac {1}{8\pi}\frac {1}{(f_L^\dagger f_L)_{ii}+(f_R^\dagger f_R)_{ii}+(\mathcal{Y}_L^\dagger \mathcal{Y}_L)_{ii}}\sum_{k\neq i}\left (  \frac {M_{E_i}^2}{M_{E_i}^2-M_{E_k}^2}\mathrm{Im}\left [ (f_R^\dagger f_R)_{ki}(\mathcal{Y}_L^\dagger \mathcal{Y}_L)_{ik} \right ]\right ),\\
\epsilon^i_{e_R^+\phi_R^{--}}&=\frac {1}{8\pi}\frac {1}{(f_L^\dagger f_L)_{ii}+(f_R^\dagger f_R)_{ii}+(\mathcal{Y}_L^\dagger \mathcal{Y}_L)_{ii}}\sum_{k\neq i}\left (  \frac {M_{E_i}^2}{M_{E_i}^2-M_{E_k}^2}\mathrm{Im}\left [ (f_R^\dagger f_R)_{ki}(\mathcal{Y}_L^\dagger \mathcal{Y}_L)_{ik} \right ]\right ),\\
\epsilon^i_{\psi_L\bar{\chi_L}}&=\frac {1}{8\pi}\frac {2}{(f_L^\dagger f_L)_{ii}+(f_R^\dagger f_R)_{ii}+(\mathcal{Y}_L^\dagger \mathcal{Y}_L)_{ii}}\sum_{k\neq i}\left (  \frac {M_{E_i}^2}{M_{E_i}^2-M_{E_k}^2}\mathrm{Im}\left [ (\mathcal{Y}_L^\dagger \mathcal{Y}_L)_{ki}(f_R^\dagger f_R)_{ik} \right ]\right ).
\end{align}

The Boltzmann equations which track the evolution of the asymmetries and the number densities are given by~\cite{Cerdeno:2006ha, Earl:2019wjw}:{\footnote{In this leptogenesis mechanism, we are mainly interested in $\nu_R$ asymmetry and neglect flavor effects, as charged lepton interactions play only a subdominant role.}} 
\begin{align}
%%%%%%%%%%%%%%%%%%%%%%%%%%%%%%%%%%%%%%%%%%%
sHz\frac {dY_{\Sigma E_i}}{dz}=&-\gamma_{i}\left ( \frac {Y_{\Sigma {E_i}}}{Y_{E_i}^{\mathrm{eq}}}-2\left \{1+\left [  \frac {Y_{\Sigma \nu_R}}{2Y_{\nu_R}^{\mathrm{eq}}}-1 \right ]\mathrm{Br}^i_{\bar{\nu}_R\phi_R^-} \right \} \right ) \nonumber\\
&-\left( \frac {Y_{\sum E_i}^2}{{4Y_{E_i}^{\mathrm{eq}}}^2} -1\right)\left\{ \gamma_{E\bar{E} \to f \bar{f}}+\gamma_{E\bar{E} \to \chi_L \bar{\chi}_L}+\gamma_{E\bar{E} \to BB} \right\},\\
\label{eq:Boltsigma_E}
%%%%%%%%%%%%%%%%%%%%%%%%%%%%%%%%%%%%%%%%%%%%
sHz\frac {dY_{\Delta E_i}}{dz}=&-\epsilon^i_{\bar{\nu}_R\phi_R^-}\gamma_i\left ( \frac {Y_{\Sigma \nu_R}}{Y_{\nu_R}^{\mathrm{eq}}}-2\right )+\gamma_i\mathrm{Br}^i_{\bar{\nu}_R\phi_R^-}\left ( -\frac {Y_{\Delta E_i}}{Y_{E_i}^{\mathrm{eq}}}-\frac {Y_{\Delta \nu_R}}{Y_{\nu_R}^{\mathrm{eq}}}+  \frac {Y_{\Sigma \nu_R}}{2Y_{\nu_R}^{\mathrm{eq}}} \frac {Y_{\Delta \phi_R^-}}{Y_{\phi_R^-}^{\mathrm{eq}}} \right ) \nonumber\\
&+\gamma_i\mathrm{Br}^i_{{e}^+_R\phi_R^{--}}\left ( -\frac {Y_{\Delta E_i}}{Y_{E_i}^{\mathrm{eq}}}-\frac {Y_{\Delta e^{-}_R}}{Y_{e_R}^{\mathrm{eq}}}+  \frac {Y_{\Delta \phi_R^{--}}}{Y_{\phi_R^{--}}^{\mathrm{eq}}} \right )+\gamma_i\mathrm{Br}^i_{{\psi}_L\bar{\chi}_L}\left ( -\frac {Y_{\Delta E_i}}{Y_{E_i}^{\mathrm{eq}}}+  \frac {Y_{\Delta \psi_L}}{Y_{\psi_L}^{\mathrm{eq}}} \right ) \nonumber\\
&+\gamma_i\mathrm{Br}^i_{{\psi}_L\phi_L}\left ( -\frac {Y_{\Delta E_i}}{Y_{E_i}^{\mathrm{eq}}}-\frac {Y_{\Delta \psi_L}}{Y_{\psi_L}^{\mathrm{eq}}}+  \frac {Y_{\Delta \phi_L}}{Y_{\phi_L}^{\mathrm{eq}}} \right )
\\
%%%%%%%%%%%%%%%%%%%%%%%%%%%%%%%%%%%%%%%%%%%
sHz\frac {dY_{\Sigma  \nu_{R}}}{dz}=& \sum_i\left\{ \gamma_i\mathrm{Br}^i_{\bar{\nu}_R\phi_R^-}\left ( \frac {Y_{\Sigma {E_1}}}{Y_{E_1}^{\mathrm{eq}}}-\frac {Y_{\Sigma {\nu_R}}}{Y_{\nu_R}^{\mathrm{eq}}} \right ) \right\}\\
%%%%%%%%%%%%%%%%%%%%%%%%%%%%%%%%%%%%%%%%%%%
sHz\frac {dY_{\Delta  \nu_{R}}}{dz}=&\sum_i\left\{-\epsilon^i_{\bar{\nu}_R\phi_R^-}\gamma_{i}\left ( \frac {Y_{\Sigma {E_i}}}{Y_{E_i}^{\mathrm{eq}}}-2\left \{1+\left [  \frac {Y_{\Sigma \nu_R}}{2Y_{\nu_R}^{\mathrm{eq}}}-1 \right ]\mathrm{Br}^i_{\bar{\nu}_R\phi_R^-} \right \} \right ) \right. \nonumber\\
& \left. -\gamma_i\mathrm{Br}^i_{\bar{\nu}_R\phi_R^-}\left (\frac {Y_{\Delta \nu_{R}}}{Y_{\nu_R}^{\mathrm{eq}}} -\frac {Y_{\Sigma \nu_R}Y_{\Delta \phi_R^-}}{2Y_{\nu_R}^{\mathrm{eq}}Y_{\phi_R^-}^{\mathrm{eq}}}+\frac {Y_{\Delta E_i}}{Y_{E_i}^{\mathrm{eq}}}  \right ) \right\} \\
%%%%%%%%%%%%%%%%%%%%%%%%%%%%%%%%%%%%%%%%%%%
sHz\frac {dY_{\Delta  \phi_{R}^-}}{dz}=&-sHz\frac {dY_{\Delta  \nu_{R}}}{dz} \\
%%%%%%%%%%%%%%%%%%%%%%%%%%%%%%%%%%%%%%%%%%%%
sHz\frac {dY_{\Delta  e_{R}^-}}{dz}=&\sum_i \left\{ -\epsilon^i_{{e}_R^+\phi_R^{--}}\gamma_{i}\left ( \frac {Y_{\Sigma {E_i}}}{Y_{E_i}^{\mathrm{eq}}}-2\left \{1+\left [  \frac {Y_{\Sigma \nu_R}}{2Y_{\nu_R}^{\mathrm{eq}}}-1 \right ]\mathrm{Br}^i_{\bar{\nu}_R\phi_R^-} \right \} \right ) \right. \nonumber \\
&\left. -\gamma_i\mathrm{Br}^i_{{e}_R^+\phi_R^{--}}\left (\frac {Y_{\Delta e^-_{R}}}{Y_{e^-_R}^{\mathrm{eq}}} -\frac {Y_{\Delta \phi_R^{--}}}{Y_{\phi_R^{--}}^{\mathrm{eq}}}+\frac {Y_{\Delta E_i}}{Y_{E_i}^{\mathrm{eq}}} 
\right ) \right\}\\
%%%%%%%%%%%%%%%%%%%%%%%%%%%%%%%%%%%%%%%%%%%%
sHz\frac {dY_{\Delta  \phi_{R}^{--}}}{dz}=&-sHz\frac {dY_{\Delta  e_{R}^-}}{dz} \\
%%%%%%%%%%%%%%%%%%%%%%%%%%%%%%%%%%%%%%%%%%%%
sHz\frac {dY_{\Delta  \psi_{L}}}{dz}=&\sum_i \left\{ \epsilon^i_{{\psi}_L\bar{\chi}_L}\gamma_{i}\left ( \frac {Y_{\Sigma {E_i}}}{Y_{E_i}^{\mathrm{eq}}}-2\left \{1+\left [  \frac {Y_{\Sigma \nu_R}}{2Y_{\nu_R}^{\mathrm{eq}}}-1 \right ]\mathrm{Br}^i_{\bar{\nu}_R\phi_R^-} \right \} \right ) \right. \nonumber\\
&\left. -\gamma_i\mathrm{Br}^i_{{\psi}_L\bar{\chi}_L}\left (\frac {Y_{\Delta \psi_L}}{Y_{\psi_L}^{\mathrm{eq}}} -\frac {Y_{\Delta E_i}}{Y_{E_i}^{\mathrm{eq}}} 
\right ) 
-\gamma_i\mathrm{Br}^i_{\bar{\psi}_L\phi_L}\left (\frac {Y_{\Delta \psi_{L}}}{Y_{\psi_L}^{\mathrm{eq}}} -\frac {Y_{\Delta \phi_L}}{Y_{\phi_L}^{\mathrm{eq}}}+\frac {Y_{\Delta E_i}}{Y_{E_i}^{\mathrm{eq}}} 
\right ) \right\}\\
%%%%%%%%%%%%%%%%%%%%%%%%%%%%%%%%%%%%%%%%%%%%
sHz\frac {dY_{\Delta  \phi_{L}}}{dz}=&\sum_i \left\{ \gamma_i\mathrm{Br}^i_{\bar{\psi}_L\phi_L}\left (\frac {Y_{\Delta \psi_{L}}}{Y_{\psi_L}^{\mathrm{eq}}} -\frac {Y_{\Delta \phi_L}}{Y_{\phi_L}^{\mathrm{eq}}}+\frac {Y_{\Delta E_i}}{Y_{E_i}^{\mathrm{eq}}} \right ) \right\}
\label{eq:Bolt}
\end{align}
where $z\equiv M_{E_1}/T $, $Y_i=n_i/s$ with $n_i$ number density of particle of $i$ , $Y_{\Sigma E_1}\equiv Y_{E}+Y_{\bar{E}}$, $Y_{\Sigma \nu_R}\equiv Y_{\nu_R}+Y_{\bar{\nu}_R}$, $s=\frac {g_{\star}2\pi^2}{45}T^3$ is the entropy density, and $H=1.66 \sqrt{g_\star}\frac {T^2}{\mathrm{Mpl}}$ is the Hubble expansion rate with $\mathrm{M_{pl}}=1.22 \times 10^{19}\,\mathrm{GeV}$. The equilibrium number density $Y_{i}^{\mathrm{eq}}$ of particles of species $i$ {with mass $m_i$} is given by,
\begin{equation}
Y_{i}^{\mathrm {eq}}=\left\{\begin{matrix}
\frac {45g_i}{4\pi^4 g_{\star}}z_i^2\mathcal{K}_2(z_i) & z_i=\frac {m_i}{T}\neq0\\
\frac {15g_i\zeta_i }{8\pi^2 g_{\star}}& m_i=0,
\end{matrix}\right. 
\end{equation}
where $\zeta_i=1(2)$ for relativistic fermions (bosons) and, $\mathcal{K}_n(x)$ are modified Bessel's function. The total decay density reaction rate of $E_i$ is given by
\begin{align}
\gamma_{i}=sY_{E_i}^{\mathrm{eq}}\Gamma_{i}\frac {\mathcal{K}_1(z)}{\mathcal{K}_2(z)}.
\end{align}
{
In~\autoref{eq:Boltsigma_E}, we have accounted for the $2 \to 2$ scattering process involving $E_i$, which can reduce the number density of $E_i$, thereby hindering the leptogenesis mechanism and the generation of $\nu_R$ asymmetry. The scattering rate density is provided by, 
\begin{equation}
\gamma_{E\bar{E} \to X \bar{X}} =\frac{T}{64\pi^4}\int_{\hat{s}_{\rm{min}}}^\infty d\hat{s} \hat{s}^{1/2} \mathcal{K}_1\left ( \frac{\sqrt{\hat{s}}}{T} \right ) \hat{\sigma}(\hat{s})_{E\bar{E} \to X \bar{X}}. 
\label{eq:scatter}
\end{equation}
Here we considered the scattering process $E_i\bar{E_i} \to f \bar{f}$ with $f$ denoting all lighter fermions, $E_i\bar{E_i} \to \chi_L \bar{\chi}_L$ and  $E_i\bar{E_i} \to B {B}$. The corresponding reduced scattering cross-sections are given by,
\begin{align}
 \hat{\sigma}(\hat{s})_{E \bar{E} \to f\bar{f} }=&\frac{119 Y^4}{288\pi}\sqrt{1-\frac{4M^2_E}{\hat{s}}}\left ( 1+\frac{2M^2_E}{\hat{s}} \right )\\
  \hat{\sigma}(\hat{s})_{E\bar{E} \to \chi_L \bar{\chi}_L}=&\frac{Y^4}{96\pi}\sqrt{1-\frac{4M^2_E}{\hat{s}}}\left ( 1+\frac{2M^2_E}{\hat{s}} \right )\\
  \hat{\sigma}(\hat{s})_{E\bar{E} \to BB}=&\frac {Y^4}{4\pi}\left[ \left( 1+\frac{4M^2_E}{\hat{s}}-\frac{16M^4_E}{\hat{s}^2} \right)\ln\left[ \frac {1+\sqrt{1-\frac{4M^2_E}{\hat{s}}}}{1-\sqrt{1-\frac{4M^2_E}{\hat{s}}}} \right]+2\left( 1+\frac{4M^2_E}{\hat{s}} \right)\sqrt{1-\frac{4M^2_E}{\hat{s}}} \right]
\end{align}
\begin{figure}[tb!]
\includegraphics[width=0.48\hsize]{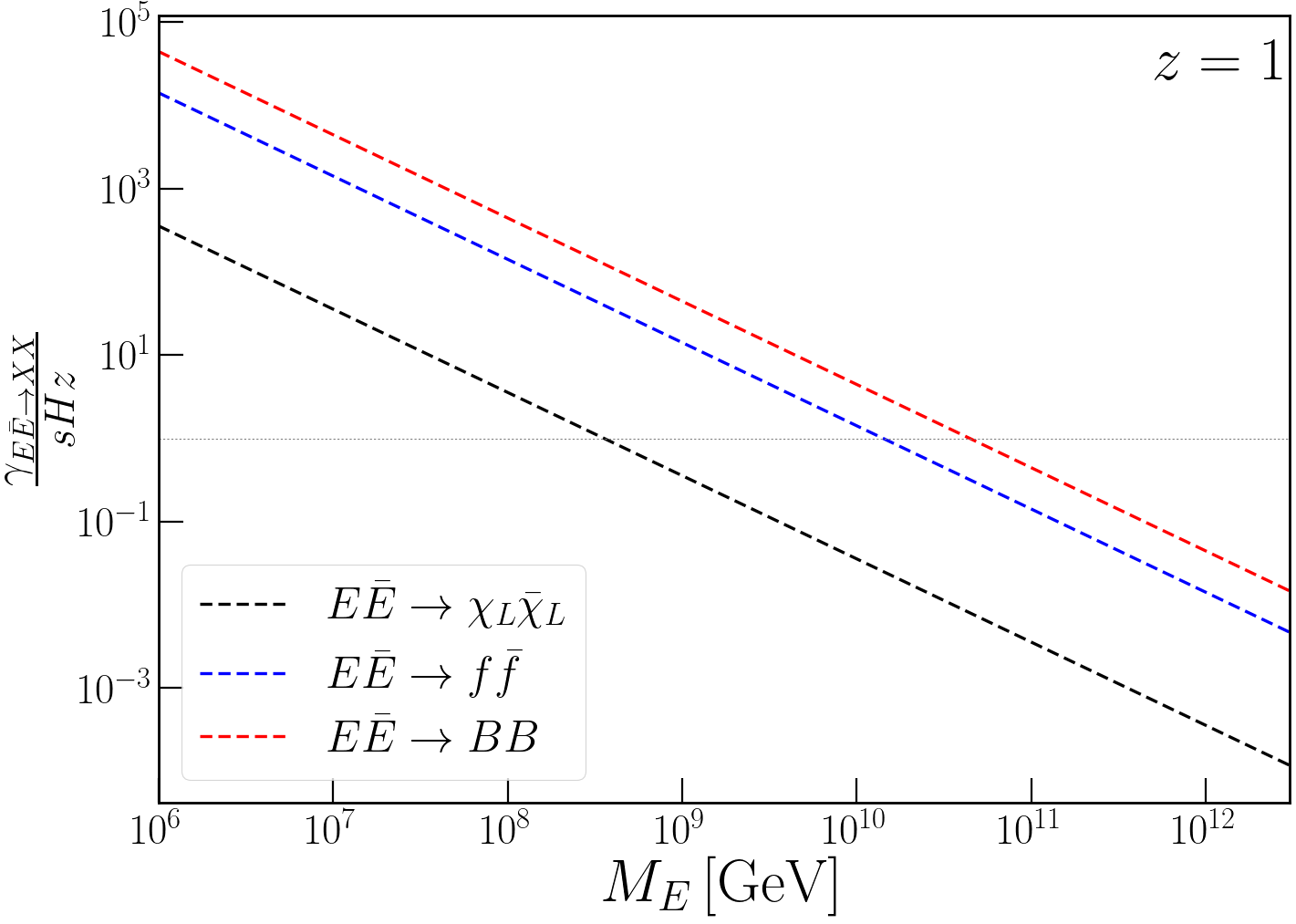}
\includegraphics[width=0.48\hsize]{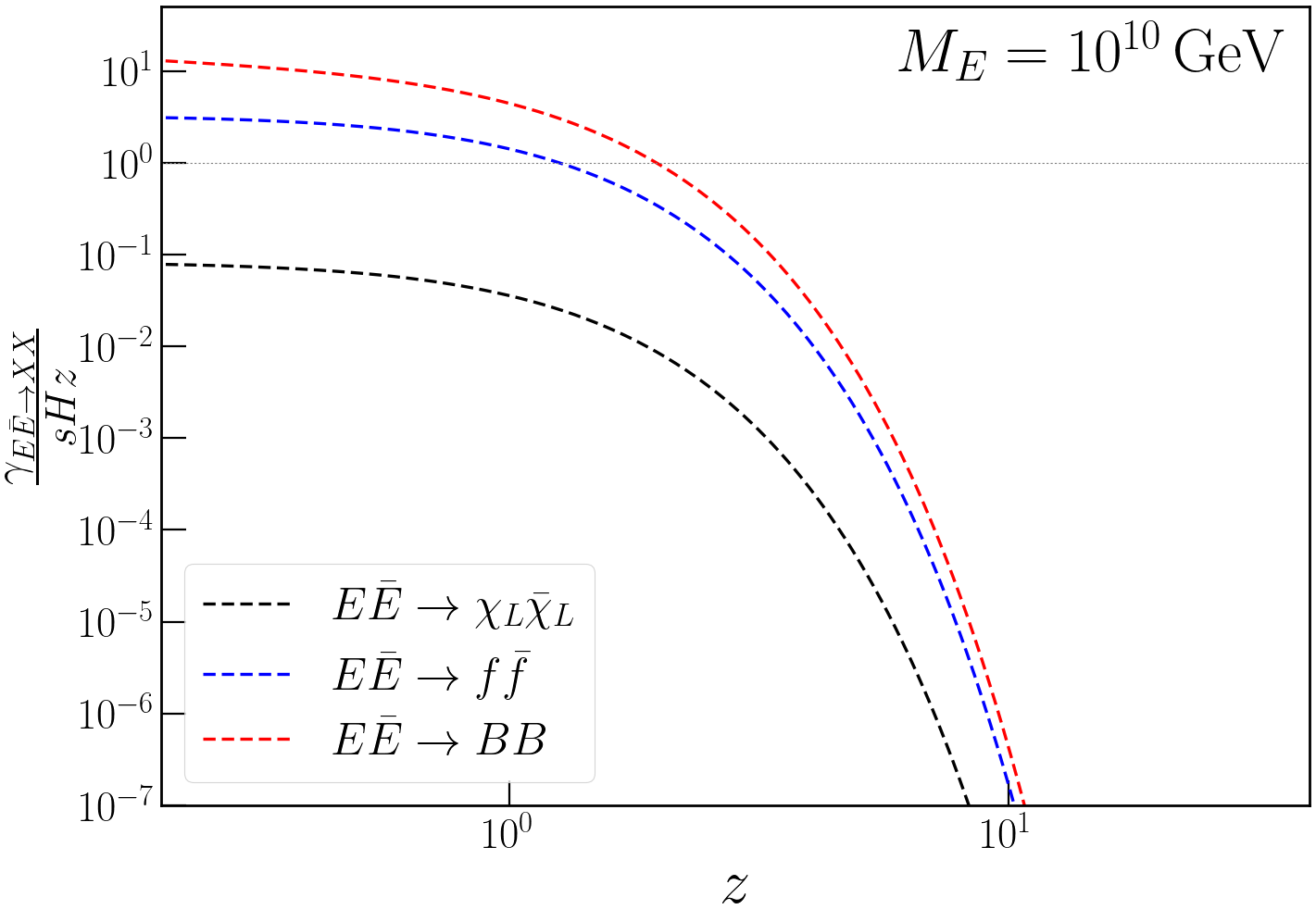}
\caption{We illustrate the relevance of the $2 \to 2$ scattering process by plotting $\gamma_{E\bar{E} \to X \bar{X}} / (sHz)$ as a function of $M_E$ (with $z = 1$ fixed) in the left panel, and as a function of $z$ (with $M_E = 10^{10}\,\mathrm{GeV}$ fixed) in the right panel. The scattering process enters thermal equilibrium at a temperature around $10^{11}\,\mathrm{GeV}$. The value of $\gamma_{E\bar{E} \to X \bar{X}} / (sHz)$ decreases with increasing $z$. 
}
\label{fig:scatterprocess}
\end{figure}
In~\autoref{fig:scatterprocess}, we examine the relevance of the $2 \to 2$ scattering process. The right panel presents $\gamma_{E\bar{E} \to X \bar{X}}/sH z$ as a function of $M_E$ with $z = 1$ fixed, while the left panel shows its dependence on $z$ with $M_E = 10^{10}\, \mathrm{GeV}$ fixed. Across the entire parameter space, the dominant annihilation channel is the vector-like lepton annihilation into hypercharge gauge bosons, $E \bar{E} \to B B$, followed by annihilation into the Standard Model fermion pairs. The scattering process enters thermal equilibrium at a temperature of approximately $10^{11}\,\mathrm{GeV}$ and becomes increasingly relevant as the Universe cools. For a fixed value of $M_E$, the quantity $\gamma_{E\bar{E} \to X \bar{X}} / (sHz)$ decreases with increasing $z$, primarily due to the suppression induced by the $\mathcal{K}_1$ term in~\autoref{eq:scatter}.

}
Following the decay of the heavy vector-like lepton,  an equal and opposite asymmetry develops in $\nu_R$ and $\phi_R^-$. Once the $\nu_R$ asymmetry is generated, the right-handed neutrinos decouple from the Standard Model plasma, provided the $\phi_R^-$ decay involving $\nu_R$ is out-of-equilibrium.
While the decay of $\phi_R^-$ into $\nu_R$ cancels the $\nu_R$ asymmetry, the two-body decay $\phi_R^-\rightarrow \phi_L \bar{\chi}_L$ preserves it. Subsequently, $\phi_L$ decays into left-handed leptons, transferring the $\phi_R^-$ asymmetry to the SM leptons. The width for the decay of $\phi_L^-$ and $\phi_R^-$ are given by:
\begin{align}
\Gamma(\phi_R^-\rightarrow {\nu}_R \psi_L \bar{\chi}_L)&=\sum_i\left ( \frac {\sum_{\alpha \beta}\left | f_{\alpha i} \right |^2\left | {\mathcal{Y}_\ell}_{\beta i} \right |^2}{32\pi^3}\frac {11}{1920}\frac {M^5_{\phi_R^-}}{M^4_{E_i}} \right ),\\
\Gamma(\phi_R^-\rightarrow\phi_L\bar\chi_L)&=\frac {\lambda_5^2\kappa_R^2}{8\pi^2M_{\phi_R^-}}\left ( 1-\frac {M^2_{\phi_L^-}}{M^2_{\phi_R^-}} \right ), \\
\Gamma(\phi_L \rightarrow \nu_L\psi_L\bar\chi_L)&=\sum_i\left ( \frac {\sum_{\alpha \beta}\left | f_{\alpha i} \right |^2\left | {\mathcal{Y}_\ell}_{\beta i} \right |^2}{64\pi^3}\frac {5}{192}\frac {M^3_{\phi_L}}{M^2_{E_i}} \right ), \\
\Gamma\left( \phi_R^- \to e_R^-+\nu_R \right)&=\frac {\mathrm{Tr}\left[ f\rho_R^\dagger \rho_R f^\dagger \right]}{8\pi}M_{\phi_R^-}~~.
\end{align}
\begin{figure}[th!]
\centering
\includegraphics[width=0.7\textwidth]{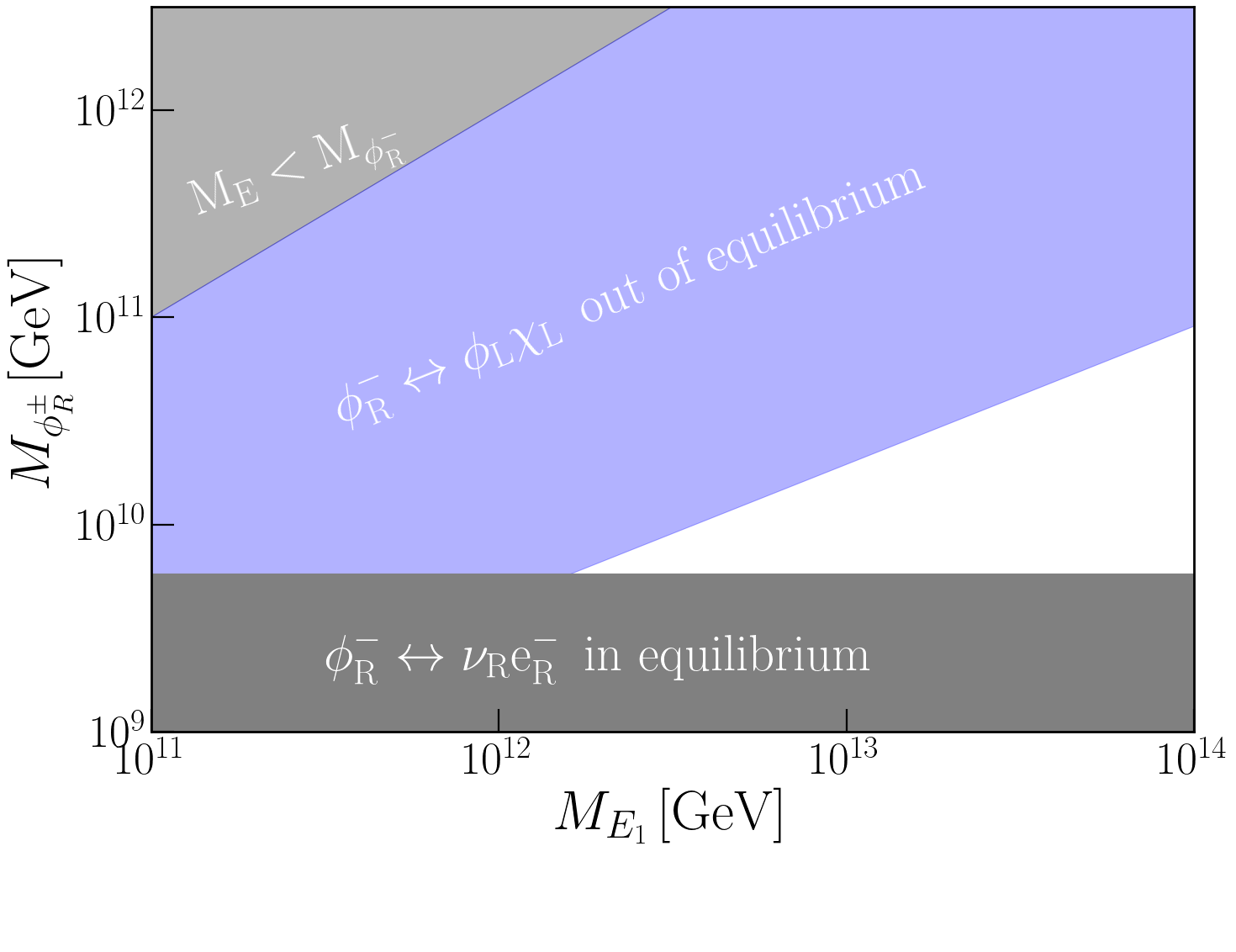}
\caption{We show the washout process constraints in the $M_E$ versus $M_{\phi_R^\pm}$ plane, assuming Yukawa coupling elements of $f$ is order $f \sim 10^{-3}$ and $M_E = 0.5 \kappa_R$. We require the process $\phi_R^-\rightarrow \phi_L^- \bar{\chi}_L^0$ to be in equilibrium and the process $\phi_R^-\rightarrow e_R^- \nu_R$ to be out of equilibrium. Additionally, to satisfy the perturbative  unitarity condition, we require that the cubic coupling, arising from $\lambda_5$ when $\kappa_R$ acquires a non-zero VEV, should not be much larger than the largest charged scalar masses $M_{\phi_R^\pm}$.}
\label{fig:const1}
\end{figure}
In the parameter space of interest, the three-body decay is typically out of equilibrium. The requirements of the process $\phi_R^-\rightarrow \phi_L \bar{\chi}_L$  be in equilibrium and $\phi_R^-\rightarrow e_R^- \nu_R$ be in out-of-equilibrium provide stringent constraints to the model parameters. In the~\autoref{fig:const1}, we illustrate these constraints in the $M_{E_1}$ versus $M_{\phi_R^\pm}$ plane, by assuming $M_{\phi_R^\pm} \approx M_{h_2^\pm}$, $M_E = 0.5 \kappa_R$ and the elements of Yukawa coupling $f$ are of order $10^{-3}$. In addition to washout conditions, the pertubative unitarity condition requires the cubic coupling arising from $\lambda_5$ when $\kappa_R$ acquires a non-zero VEV, should not be much larger than the largest charged scalar masses and here we require it to be less than $3M_{\phi_R^\pm}$. These conditions restrict the mass of heavy Dirac leptons to be roughly $M_{E}>10^{12}\,\mathrm{GeV}$. The mass of $E_i$ can be lowered down in the case of degenerate heavy leptons, utilizing the resonant enhancement of CP asymmetry. The resonant leptogenesis scenario will be discussed in more detail in~\autoref{sec:res}. For a given $M_E$, the upper bound on $M_{\phi_R^-}$ is set by the condition that the process $\phi_R^-\rightarrow \phi_L \bar{\chi}_L$ remains in equilibrium, while the lower bound is obtained by ensuring that $\phi_R^-\rightarrow e_R^- \nu_R$ remains out of equilibrium. Additionally, we require the decay rate $\Gamma(\phi_L \rightarrow \psi_L\psi_L\bar\chi_L)> \left. H(T)  \right|_{T=T_{\mathrm{sph}}}$, where $T_{\mathrm{sph}}=131.7\pm 2.3 \,\mathrm{GeV}$~\cite{DOnofrio:2014rug} is the sphaleron decoupling temperature. This condition ensures the asymmetry in $\phi_L$ is efficiently transferred to SM leptons before electroweak symmetry breaking.

In this model, $B-L$ is conserved, and we have the asymmetry $Y_{\Delta (B-L_{\mathrm{SM}})}=Y_{\Delta \nu_R}$. The electroweak sphaleron will convert $B-L_{\mathrm{SM}}$ asymmetry into baryon asymmetry~\cite{PhysRevD.42.3344, Kuzmin:1985mm},
\begin{equation}
Y_{\Delta B}=\frac {28}{79}Y_{\Delta (B-L_{\mathrm {SM}})}=\frac {28}{79} Y_{\Delta \nu_R}.
\end{equation}
The observed baryon to entropy ratio is given by $Y_{\Delta B}=(8.579\pm0.109)\times 10^{-11}$~\cite{Planck:2018vyg}.

%%%%%%%%%%%%%%%%%%%%%%%%%%%%%%%%%%%%%%%%%%%%%%%%%%%%%%%%%%%%
\subsection{Results}
\begin{table}
\centering
\begin{tabular}{
|c||c|c|c|}
\hline
\multicolumn{4}{|c|}{Benchmark Points} \\
\hline
& BP1  & BP2 & BP3  \\
\hline
\hline
$M_E\,[\mathrm{GeV}]$& $\mathrm{diag}\left( 3,6,9 \right)\times 10^{12}$& $\mathrm{diag}\left( 3,7,11 \right)\times 10^{12}$&$\mathrm{diag}\left( 4,6,11 \right)\times 10^{12}$\\
\hline
$\theta_{\pm}$&  $4.2727\times 10^{-10}$& $4.7396\times 10^{-10}$ & $-7.9305\times 10^{-10}$ \\
\hline
$\sin^2\theta_{12}$& $0.3066$& $0.3057$&$0.3080$ \\
\hline
$\sin^2\theta_{13}$& $0.0220$& $0.0220$& $0.0222$ \\
\hline
$\sin^2\theta_{23}$& $0.5676$& $0.5913$& $0.5620$ \\
\hline
$\delta_{\mathrm{CP}}$&  $164.4^\circ$& $247.3^\circ$ & $285.0$\\
\hline
$\Delta m_{21}^2[10^{-5}\mathrm{eV}]$&  $7.4148$& $7.4121$ &$7.4900$ \\   
\hline
$\Delta m_{31}^2[10^{-3}\mathrm{eV}]$ [NO]& $2.509$& $2.5106$& -\\
$\Delta m_{23}^2[10^{-3}\mathrm{eV}]$ [IO]& -& -& $2.5100$\\
\hline
$Y_{\Delta B}$& $8.5791\times 10^{-11}$& $8.569\times 10^{-11}$& $8.5790 \times 10^{-11}$\\
\hline
$\chi^2$& $0.2565$& $0.8068$& $7.32\times 10^{-8}$\\
\hline 
\end{tabular}  
\caption{Three benchmark point fit for the neutrino oscillation data and baryon asymmetry in the model by fixing $\kappa_R=10^{13}\,\mathrm{GeV}$, $m_{h_2^\pm}=10^{9}\,\mathrm{GeV}$ and $m_{h_1^\pm}=10^{7}\,\mathrm{GeV}$. Here NO and IO refer to the normal and inverted orderings of neutrino masses.
}
\label{tab:1}
\end{table}

To fit the neutrino oscillation data and baryon asymmetry, we perform a $\chi^2$-function minimization, where the $\chi^2$-function is defined as
\begin{equation}
\chi^2=\sum_{\mathrm{all\,\, observables}} \left ( \frac{\mathrm{theoretical\,predictions}-\mathrm{\, experimental\, central\, value}}{\mathrm{experimental\,}1\sigma\mathrm{\,error}} \right )^2.
\end{equation}

The model can give excellent fits to neutrino oscillation~\cite{Esteban:2024eli} data and baryon asymmetry for approximately $\kappa_R>10^{13}\,\mathrm{GeV}$, as illustrated in~\autoref{tab:1}. This limit primarily comes from the requirement of the process $\phi_R^-\rightarrow e_R^- \nu_R$ to be out of equilibrium which requires approximately $M_{\phi_R^-}>10^{9}\,\mathrm{GeV}$. 
This constrain can be alleviated in the resonant leptogenesis regime, which allows the scale of Yukawa coupling $f$ to even lower and could keep the process $\phi_R^-\rightarrow e_R^- \nu_R$ out of equilibrium for lower values of $M_{\phi_R^-}$. In~\autoref{tab:1}, we provide benchmark points by fixing $\kappa_R=10^{13}\,\mathrm{GeV}$, $m_{h_2}=10^{9}\,\mathrm{GeV}$ and $m_{h_1}=10^{7}\,\mathrm{GeV}$. Here, we assume the $M_{\phi_R^-}\approx M_{h_2}$ and $M_{\phi_L^-}\approx M_{h_1}$.  The benchmark points BP1 and BP2 correspond to normal ordering (NO), while BP3 corresponds to inverted ordering (IO) of neutrino masses. The Yukawa coupling matrices associated with the benchmark points are provided in~\autoref{sec:Yuk}.

\begin{figure}[tb!]
\includegraphics[width=0.48\hsize]{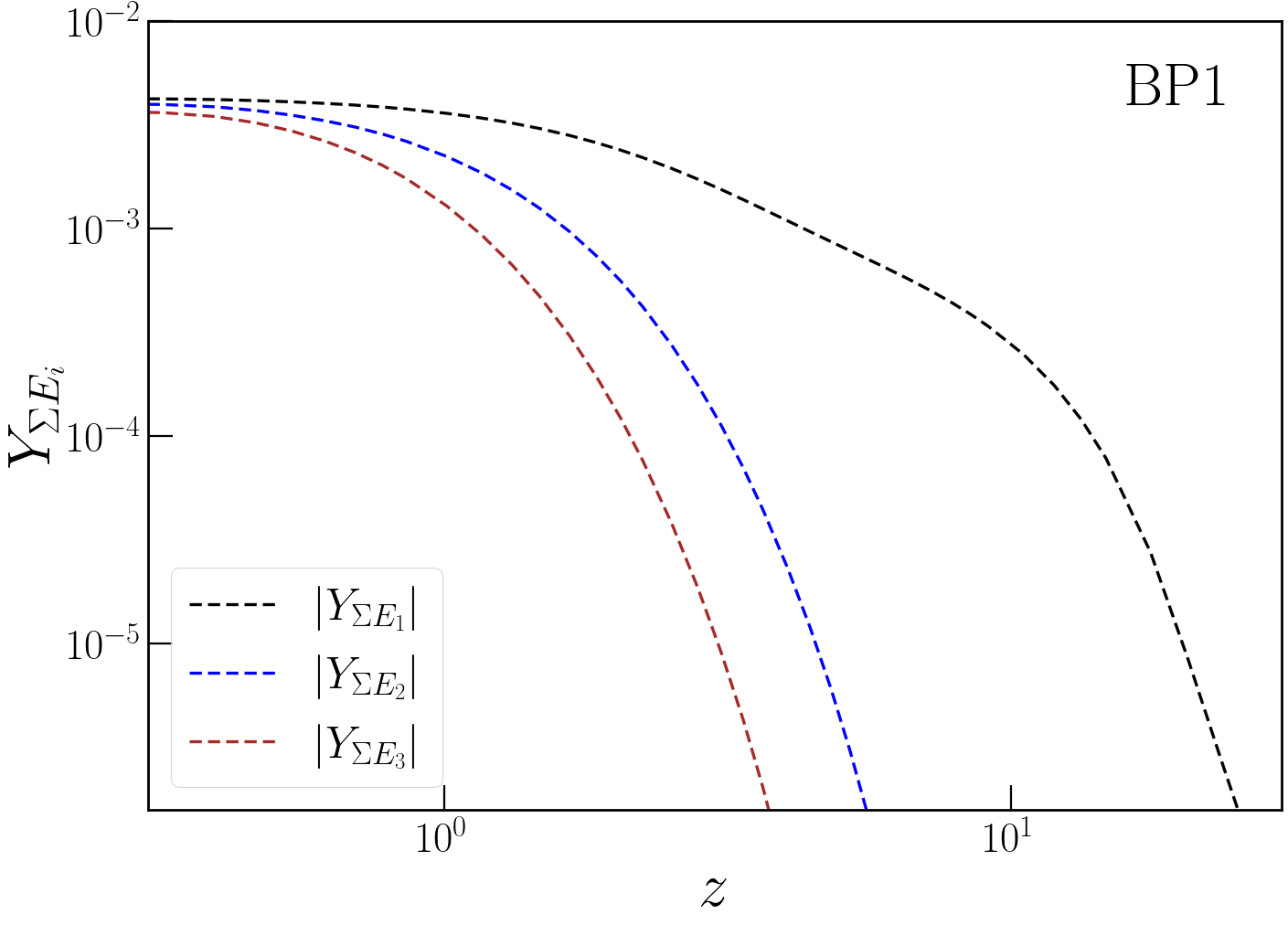}
\includegraphics[width=0.48\hsize]{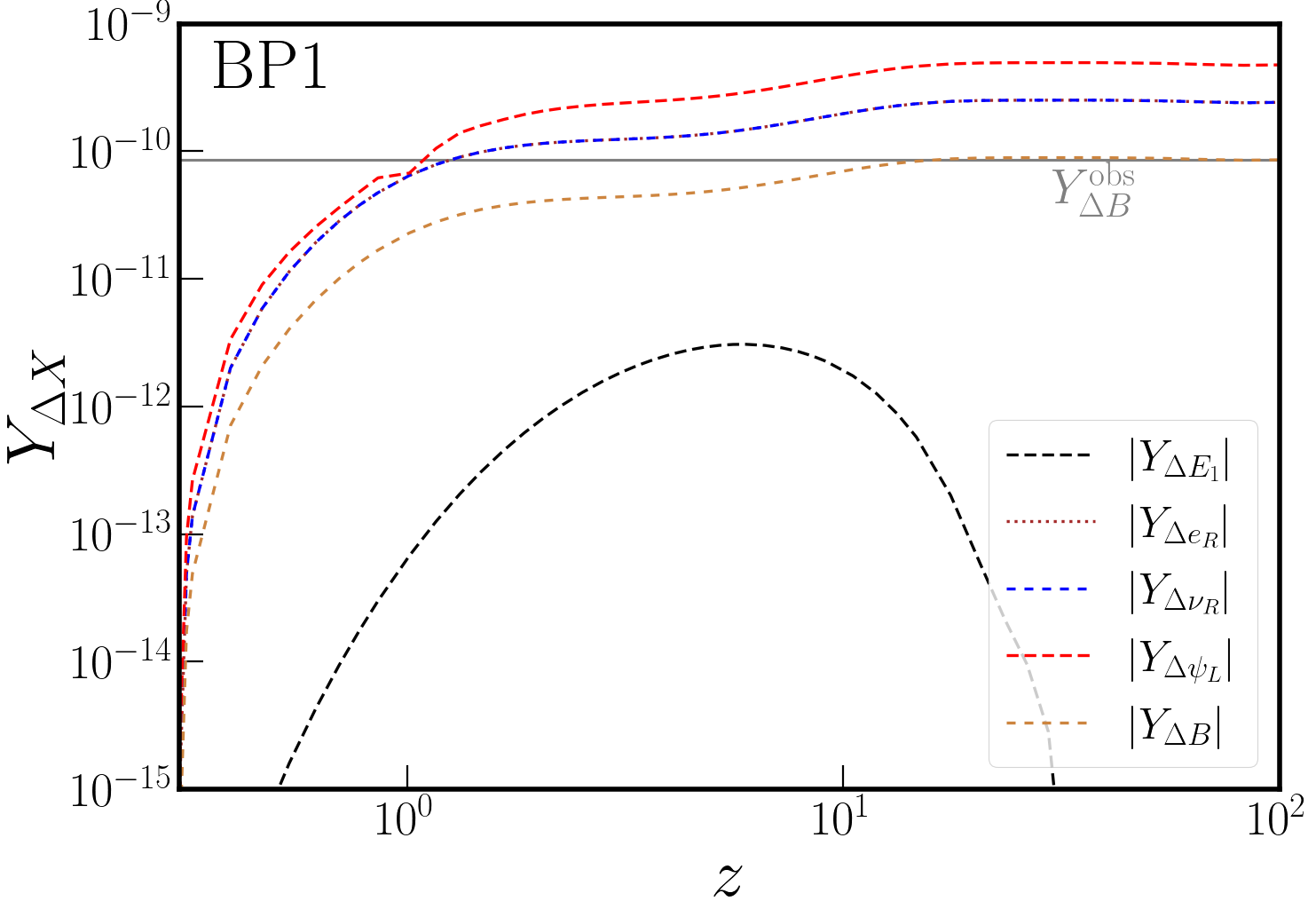}
\caption{The evolution of number densities of the three Dirac leptons $Y_{\Sigma E_i}$ (left panel) and asymmetries $Y_{\Delta \nu_R}$, $Y_{\Delta e_R}$, $Y_{\Delta \psi_L}$ and $Y_{\Delta E_1}$ (right panel) obtained by solving Boltzman equations of~\autoref{eq:Bolt}. 
}
\label{fig:BP1}
\end{figure}

In~\autoref{fig:BP1}, we present the evolution of the number densities of Dirac leptons $Y_{\Sigma E_1}\, Y_{\Sigma E_2}$ and $Y_{\Sigma E_3}$ (left panel) and the asymmetries $Y_{\Delta \nu_R}$, $Y_{\Delta e_R}$, $Y_{\Delta \psi_L}$, and $Y_{\Delta E_1}$ (right panel), obtained by solving the Boltzmann equations in~\autoref{eq:Bolt}. The asymmetries of $Y_{\Delta \nu_R}$, $Y_{\Delta e_R}$ and, $Y_{\Delta \psi_L}$ are almost of the same order because the CP asymmetries of the corresponding channels are of similar order. The asymmetry of $Y_{\Delta E_1}$ is significantly small compared to other asymmetries because the sum of CP violation of all decay channels of $E_1$ vanishes. In the weak washout regime, the asymmetries do not undergo significant washout. In this scale, the contributions from the heavier Dirac states $E_2$ and $E_3$ are negligible, as their CP violation is minimal and does not substantially impact the washout of the $\nu_R$ asymmetry. This model can give an excellent fit for neutrino oscillation data and explain the observed baryon asymmetry but with right-hand breaking scale $\kappa_R$ to be approximately above $10^{13}\, \mathrm{GeV}$. However, this scale can be lowered using resonant enhancement of CP asymmetry, as discussed in~\autoref{sec:res}.
\section{Resonant Leptogenesis}
\label{sec:res}
The CP asymmetry can be resonantly enhanced due to the mixing of nearly degenerate heavy Dirac states having mass differences similar order to their decay widths~\cite{Pilaftsis:2003gt}.  This enhancement in CP asymmetry opens a pathway to reproduce the observed baryon asymmetry with smaller values of the $f$ Yukawa coupling, thereby lowering the required scale of $\kappa_R$. In this scenario, the CP asymmetry generated due to the decay of $\hat{E}_1$ is provided by~\cite{PhysRevD.56.5431,Pilaftsis:2003gt}:
\begin{align}
\epsilon^1_{\bar{\nu_R}\phi_R^-}&=\frac {\mathrm{Im}\left [ ({f_R}^\dagger {f_R})_{21}(\mathcal{{Y}}_L^\dagger \mathcal{{Y}}_L)_{12} \right ]\Lambda\left( M_{E_1},M_{E_2},\Gamma_2 \right)}{\left[ ({f_L}^\dagger {f_L})_{11}+({f_R}^\dagger {f_R})_{11}+(\mathcal{{Y}}_L^\dagger \mathcal{{Y}}_L)_{11} \right]\left[ ({f_L}^\dagger {f_L})_{22}+({f_R}^\dagger {f_R})_{22}+(\mathcal{{Y}}_L^\dagger \mathcal{{Y}}_L)_{22} \right]},\\
\epsilon^1_{e_R^+\phi_R^{--}}&=\frac {\mathrm{Im}\left [ ({f_R}^\dagger {f_R})_{21}(\mathcal{{Y}}_L^\dagger \mathcal{{Y}}_L)_{12} \right ]\Lambda\left( M_{E_1},M_{E_2},\Gamma_2 \right)}{\left[ ({f_L}^\dagger {f_L})_{11}+({f_R}^\dagger {f_R})_{11}+(\mathcal{{Y}}_L^\dagger \mathcal{{Y}}_L)_{11} \right]\left[ ({f_L}^\dagger {f_L})_{22}+({f_R}^\dagger {f_R})_{22}+(\mathcal{{Y}}_L^\dagger \mathcal{{Y}}_L)_{22} \right]},\\
\epsilon^1_{\psi_L\bar{\chi_L}}&=\frac {2\mathrm{Im}\left [ (\mathcal{{Y}}_L^\dagger \mathcal{{Y}}_L)_{21}({f}_R^\dagger {f}_R)_{12} \right ]\Lambda\left( M_{E_1},M_{E_2},\Gamma_2 \right)}{\left[ ({f_L}^\dagger {f_L})_{11}+({f_R}^\dagger {f_R})_{11}+(\mathcal{{Y}}_L^\dagger \mathcal{{Y}}_L)_{11} \right]\left[ ({f_L}^\dagger {f_L})_{22}+({f_R}^\dagger {f_R})_{22}+(\mathcal{{Y}}_L^\dagger \mathcal{{Y}}_L)_{22} \right]},
\end{align}
where the factor $\Lambda\left( M_{E_1},M_{E_2},\Gamma_2 \right)$ is given by
\begin{equation}
\Lambda\left( M_{E_1},M_{E_2},\Gamma_2 \right)=\frac {\left (M_{E_2}^2-M_{E_1}^2  \right )M_{E_1}\Gamma_{2}}{\left (M_{E_2}^2-M_{E_1}^2  \right )^2+M_{E_1}^2\Gamma_{2}^2}
.
\label{eq:resYterm}
\end{equation}
Similarly, the CP asymmetry of $\hat{E}_2$ decay can be written by interchanging the indices $1\leftrightarrow 2$.  The Boltzmann equation in this scenario can be obtained by including of decay and washout of $\hat{E_2}$ contribution to~\autoref{eq:Bolt}.

Achieving a resonant scenario with a diagonal bare mass matrix $M_E$ may appear to be significantly fine-tuned. This conclusion is, however, not basis-invariant. In this work, we consider the bare matrix $M_E$ to be non-diagonal and for simplicity, we consider a two-family mixing scenario where the off-diagonal elements $M_{12}$ and $M_{21}$ are significantly larger than the diagonal elements $M_{11}$ and $M_{22}$ of the bare mass matrix $M_E$. Owing to parity symmetry we have $M_{12}$  = $M_{21} ^*$. This parametrization of the bare mass matrix would lead to the squared masses of the eigenvalues being nearly degenerate. The parametrization of Yukawa matrix $\mathcal{Y}_\ell$ with $\mathcal{Y}_{\ell\, 11},\mathcal{Y}_{\ell\, 12} \ll  \mathcal{Y}_{\ell\, 21},\mathcal{Y}_{\ell\, 22}$ can also can fit the electron and muon masses.

\begin{figure}[tb!]
\includegraphics[width=0.48\hsize]{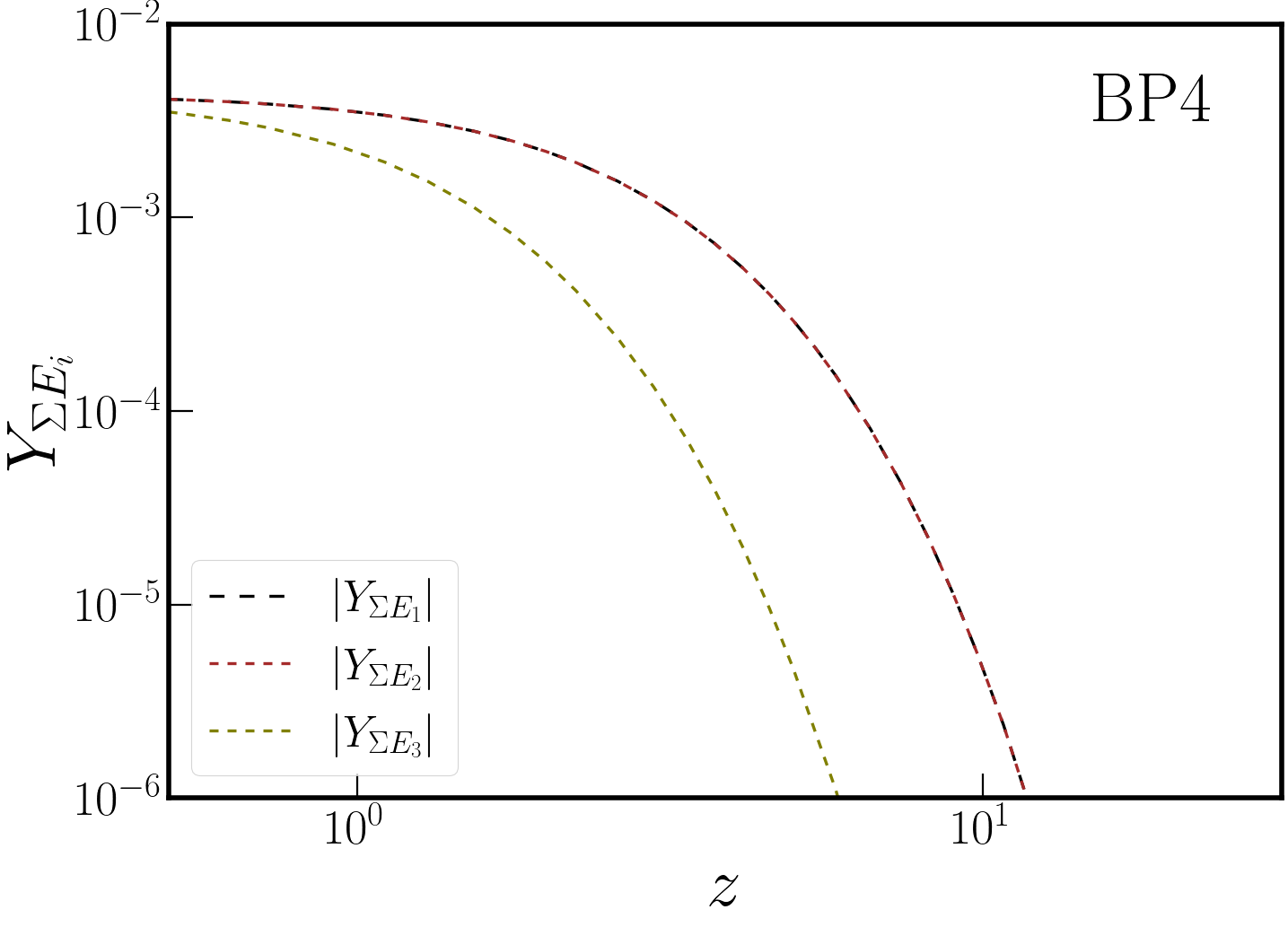}
\includegraphics[width=0.48\hsize]{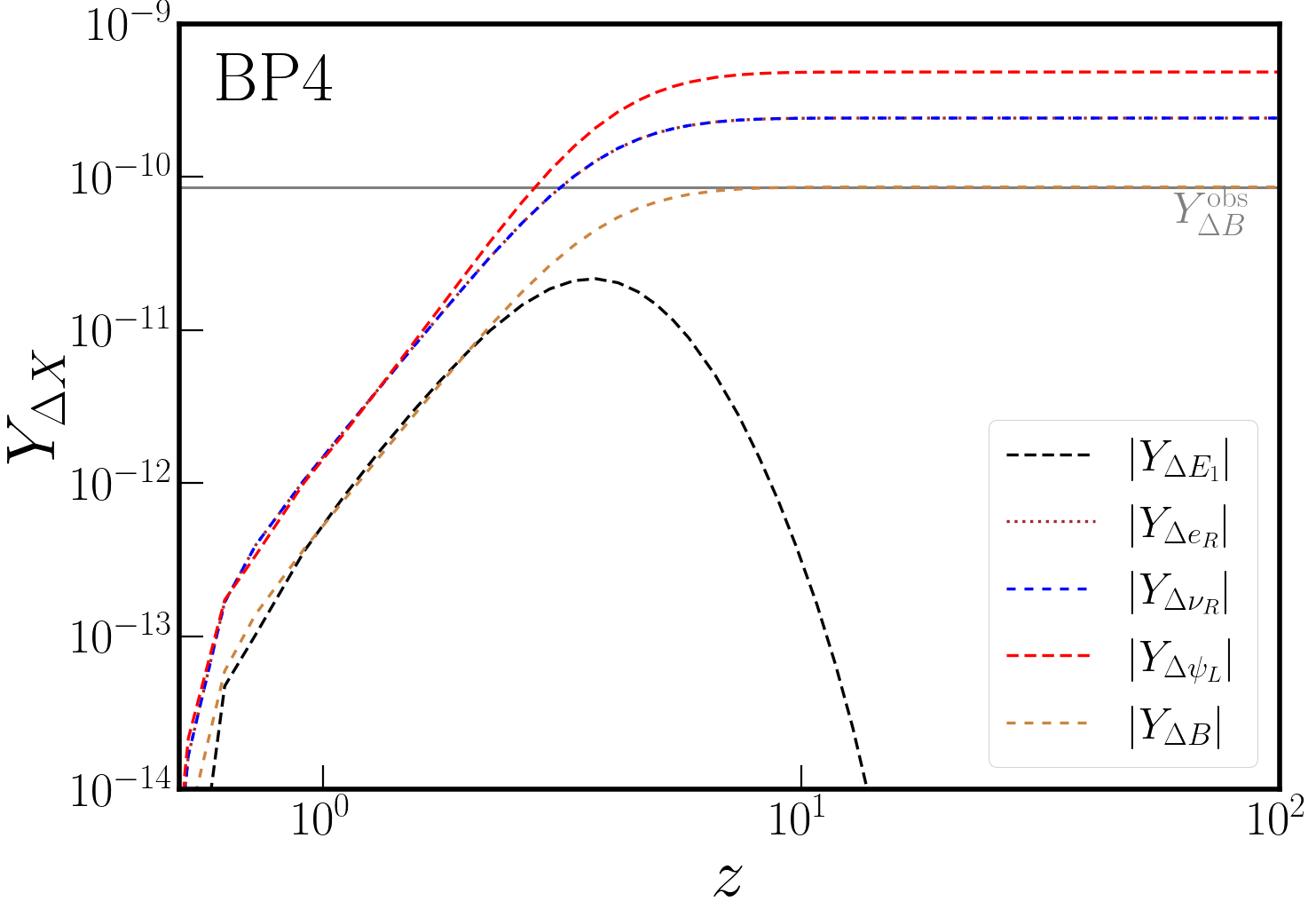}
\caption{The evolution of number densities of Dirac leptons $Y_{\Sigma E_1}$ and $Y_{\Sigma E_2}$ (left panel) and asymmetries $Y_{\Delta \nu_R}$, $Y_{\Delta e_R}$, $Y_{\Delta \psi_L}$, $Y_{\Delta E_1}$ (right panel) obtained by solving Boltzman equations of~\autoref{eq:Bolt}. }
\label{fig:BP4}
\end{figure}

\begin{table}
\centering
\begin{tabular}{
|c||c|c|c|}
\hline
\multicolumn{4}{|c|}{Resonant Leptogenesis-Benchmark Points} \\
\hline
& BP4  & BP5 & BP6  \\
\hline
\hline
$\kappa_R\,[\mathrm{GeV}]$&$10^{11}$& $10^{11}$& $10^{11}$\\
\hline
$m_{h_2^\pm}\,[\mathrm{GeV}]$&$5 \times 10^9$& $4 \times 10^9$& $3 \times 10^9$\\
\hline
$m_{h_1^\pm}\,[\mathrm{GeV}]$&$5 \times 10^8$& $4 \times 10^8$& $3 \times 10^8$\\
\hline
$\theta_{\pm}$& $2.0853 \times 10^{-9}$& $-1.5827 \times 10^{-9}$& $2.0329\times 10^{-9}$ \\
\hline
$\sin^2\theta_{12}$& $0.3070$& $0.3061$& $0.3069$ \\
\hline
$\sin^2\theta_{13}$& $0.02203$& $0.02207$& $0.02203$ \\
\hline
$\sin^2\theta_{23}$& $0.5723$& $0.5849$& $0.5721$ \\
\hline
$\delta_{\mathrm{CP}}$& $196.7^\circ$& $273.1^\circ$& $203.6^\circ$ \\
\hline
$\Delta m_{21}^2[10^{-5}\mathrm{eV}]$& $7.4097$& $7.3572$& $7.4100$ \\   
\hline
$\Delta m_{31}^2[10^{-3}\mathrm{eV}]$& $2.5113$& $2.5021$& $2.5110$\\
\hline
$Y_{\Delta B}$& $8.5789 \times 10^{-11}$& $8.5797\times 10^{-11}$& $8.5789\times 10^{-11}$\\
\hline
$\chi^2$& $0.017$& $1.502$& $0.0180$\\
\hline 
\end{tabular}  
\caption{Three benchmark points that for the neutrino oscillation data and baryon asymmetry in the resonant leptogenesis scenario. Here we have fixed $M_{E_3}=\kappa_R$.}
\label{tab:2}
\end{table}

In~\autoref{tab:2}, we provide the benchmark fit for the neutrino oscillation data and baryon asymmetry by fixing fix $M_{E_3} = \kappa_R$.{ \footnote{To simplify the numerical procedure used to identify the benchmark points in~\autoref{tab:2}, we first fit the Yukawa matrix $Y_\ell$ by maximizing~\autoref{eq:resYterm}, while ensuring consistency with the SM lepton masses. Using this fixed $Y_\ell$, we then fit the Yukawa matrix $f$ to explain the observed neutrino oscillation data and the baryon asymmetry of the Universe.}} As in the previous section, we assume $M_{\phi_R^-} \approx M_{h_2}$ and $M_{\phi_L^-} \approx M_{h_1}$.
{ In this scenario, the right-handed breaking scale $\kappa_R$ can be lowered to approximately $10^{11}\, \mathrm{GeV}$. The scattering process $ E\bar {E} \to X\bar {X}$ primarily constrains any further lowering of the $\kappa_R$ scale. The Yukawa coupling matrices and bare mass matrices for the benchmark points in~\autoref{tab:2} are provided in~\autoref{sec:Yuk}. In~\autoref{fig:BP4}, we provide the evolution of asymmetries and number density of heavy Dirac state $E_i$ as a function of $z=\hat{M}_{E_1}/T$ for the benchmark point BP4.  Hence, the resonant enhancement of asymmetry allows a large parameter space to simultaneously explain baryon asymmetry and neutrino oscillation data in the model with $SU(2)_R$ breaking scale $\kappa_R$ as low as $10^{11}\,\mathrm{GeV}$. }
%%%%%%%%%%%%%%%%%%%%%%%%%%%%%%%%%%%%%%%%%%%%%%%%%%%%%%%%%%%%%%%%%%%%
\section{Summary}
\label{sec:summ}

We have presented in this paper a novel way of generating baryon asymmetry of the Universe via Dirac leptogenesis.  The model presented is based on the left-right symmetric gauge group with a generalized seesaw mechanism inducing all charged fermion masses.  Thus, the electron mass arises via mixing with a vector-like electron ($E$) field.  Neutrinos are naturally Dirac fermions with small masses, since they do not have vector-like partners to mix with and therefore do not acquire tree-level masses. Loop diagrams involving charged scalars and the vector-like lepton field will induce finite and small Dirac neutrino masses. The decay of $E \rightarrow \overline{\nu}_R \phi_R^-$ generates a $\nu_R$ asymmetry at early times in the evolution of the Universe, which is accompanied by an equal and opposite $\phi_R^-$ asymmetry.  This $\phi_R^-$ asymmetry is converted in a second stage to $\phi_L^-$ asymmetry which in turn is transferred to standard model lepton asymmetry.

We have identified parameter space of the model where Dirac leptogenesis and excellent fits to neutrino oscillation data are simultaneously reproduced.  Typically this requires the $W_R^\pm$ gauge boson mass to exceed $10^{13}$ GeV.  However, we have also shown that a resonant enhancement is possible for the CP asymmetry quite naturally, in which case the $W_R^\pm$ mass can be as low as $10^{11}$ GeV.  
%Since the model setup is motivated independently from a solution to the strong CP problem via parity symmetry without needing the axion, it is interesting that the low $W_R^\pm$ scale can also protect the parity solution from Planck suppressed operators.

%%%%%%%%%%%%%%%%%%%%%%%%%%%%%%%%%%%%%%%%%%%%%%%%%%%%%%%%%%%%%%%%%%%%
\section*{Acknowledgments}
%%%%%%%%%%%%%%%%%%%%%%%%%%%%%%%%%%%%%%%%%%%%%%%%%%%%%%%%%%%%%%%%%%%%
%%%%%%%%%%%%%%%%%%%%%%%%%%%%%%%%%%%%%%%%%%%%%%%%%%%%%%%%%%%%%%%%%%%%
The work of KSB and AK are supported in part by US Department of Energy Grant Number DE-SC 0016013.  We thank Ritu Dcruz and Rabi Mohapatra for helpful discussions.

%%%%%%%%%%%%%%%%%%%%%%%%%%%%%%%%%%%%%%%%%%%%
%%%%%%%%%%%%%%%%%%%%%%%%%%%%%%%%%%%%%%%%%%%%%%%%%%%%%%%%%%%
\appendix
\section{Yukawa Matrices for benchmark points}

\label{sec:Yuk}
\subsection{Non-resonant leptogenesis}
In this section, we provide the Yukawa matrices for the benchmark points given in~\autoref{tab:1}.
\begin{itemize}
\item Benchmark point 1
\begin{align}
\mathcal{Y}_\ell =& 
\begin{pmatrix}
\phantom{+}9.3808 \times 10^{-4} & \hspace{15pt} \phantom{+}1.1101 \times 10^{-5} & \hspace{15pt} \phantom{+}9.0359 \times 10^{-6} \\
\phantom{+}2.6782 \times 10^{-4} & \hspace{15pt} \phantom{+}3.0142 \times 10^{-2} & \hspace{15pt} \phantom{+}1.1207 \times 10^{-2} \\
\phantom{+}4.8019 \times 10^{-4} & \hspace{15pt} \phantom{+}4.8570 \times 10^{-2} & \hspace{15pt} \phantom{+}3.2374 \times 10^{-2}
\end{pmatrix} \nonumber \\
+ i& 
\begin{pmatrix}
\phantom{+}1.5804 \times 10^{-7} & \hspace{15pt} \phantom{+}2.8662 \times 10^{-5} & \hspace{15pt} \phantom{+}2.4682 \times 10^{-5} \\
\phantom{+}3.7411 \times 10^{-4} & \hspace{15pt} -1.3378 \times 10^{-2} & \hspace{15pt} -2.9516 \times 10^{-2} \\
\phantom{+}6.6206 \times 10^{-4} & \hspace{15pt} -3.9877 \times 10^{-2} & \hspace{15pt} -1.1349 \times 10^{-2}
\end{pmatrix} \\
f =& 
\begin{pmatrix}
\phantom{+}3.6224 \times 10^{-4} & \hspace{15pt} \phantom{+}2.4159 \times 10^{-6} & \hspace{15pt} -5.5434 \times 10^{-4} \\
-2.6873 \times 10^{-4} & \hspace{15pt} -5.8284 \times 10^{-5} & \hspace{15pt} \phantom{+}2.8690 \times 10^{-4} \\
-2.6881 \times 10^{-4} & \hspace{15pt} -2.0463 \times 10^{-4} & \hspace{15pt} \phantom{+}1.0009 \times 10^{-3}
\end{pmatrix} \nonumber \\
+ i& 
\begin{pmatrix}
-1.0050 \times 10^{-4} & \hspace{15pt} -3.8727 \times 10^{-4} & \hspace{15pt} \phantom{+}5.1962 \times 10^{-5} \\
-1.0703 \times 10^{-3} & \hspace{15pt} \phantom{+}1.4418 \times 10^{-3} & \hspace{15pt} -1.4815 \times 10^{-3} \\
-4.1369 \times 10^{-4} & \hspace{15pt} -1.5804 \times 10^{-4} & \hspace{15pt} \phantom{+}1.4642 \times 10^{-5}
\end{pmatrix}
\end{align}

\item Benchmark point 2
\begin{align}
\mathcal{Y}_\ell =& 
\begin{pmatrix}
-9.3797 \times 10^{-4} & \hspace{15pt} -1.6271 \times 10^{-5} & \hspace{15pt} -1.1164 \times 10^{-5} \\
\phantom{+}3.3239 \times 10^{-4} & \hspace{15pt} \phantom{+}1.7466 \times 10^{-2} & \hspace{15pt} \phantom{+}6.9582 \times 10^{-3} \\
\phantom{+}7.3119 \times 10^{-4} & \hspace{15pt} \phantom{+}3.5579 \times 10^{-2} & \hspace{15pt} \phantom{+}2.2248 \times 10^{-2}
\end{pmatrix} \nonumber \\
+ i& 
\begin{pmatrix}
-2.0532 \times 10^{-7} & \hspace{15pt} -3.8928 \times 10^{-5} & \hspace{15pt} -3.5515 \times 10^{-5} \\
\phantom{+}2.0232 \times 10^{-4} & \hspace{15pt} -1.9426 \times 10^{-2} & \hspace{15pt} -3.7042 \times 10^{-2} \\
\phantom{+}4.4132 \times 10^{-4} & \hspace{15pt} -6.3230 \times 10^{-2} & \hspace{15pt} -2.5120 \times 10^{-2}
\end{pmatrix} \\
f =& 
\begin{pmatrix}
\phantom{+}2.1920 \times 10^{-4} & \hspace{15pt} -1.7829 \times 10^{-6} & \hspace{15pt} -2.9198 \times 10^{-4} \\
\phantom{+}4.3409 \times 10^{-4} & \hspace{15pt} -5.2722 \times 10^{-4} & \hspace{15pt} \phantom{+}3.4808 \times 10^{-4} \\
-6.9206 \times 10^{-4} & \hspace{15pt} \phantom{+}1.1962 \times 10^{-3} & \hspace{15pt} -9.7546 \times 10^{-4}
\end{pmatrix} \nonumber \\
+ i& 
\begin{pmatrix}
\phantom{+}1.0191 \times 10^{-5} & \hspace{15pt} \phantom{+}1.7818 \times 10^{-4} & \hspace{15pt} -6.7537 \times 10^{-4} \\
-6.8863 \times 10^{-5} & \hspace{15pt} -4.7936 \times 10^{-4} & \hspace{15pt} \phantom{+}1.1545 \times 10^{-3} \\
-8.1259 \times 10^{-4} & \hspace{15pt} \phantom{+}8.1527 \times 10^{-4} & \hspace{15pt} -6.1894 \times 10^{-4}
\end{pmatrix}
\end{align}

\item Benchmark point 3
\begin{align}
\mathcal{Y}_\ell =& 
\begin{pmatrix}
\phantom{+}9.3808 \times 10^{-4} & \hspace{15pt} \phantom{+}1.1101 \times 10^{-5} & \hspace{15pt} \phantom{+}9.0359 \times 10^{-6} \\
\phantom{+}2.6782 \times 10^{-4} & \hspace{15pt} \phantom{+}3.0142 \times 10^{-2} & \hspace{15pt} \phantom{+}1.1207 \times 10^{-2} \\
\phantom{+}4.8019 \times 10^{-4} & \hspace{15pt} \phantom{+}4.8570 \times 10^{-2} & \hspace{15pt} \phantom{+}3.2374 \times 10^{-2}
\end{pmatrix} \nonumber \\
+ i& 
\begin{pmatrix}
\phantom{+}1.5804 \times 10^{-7} & \hspace{15pt} \phantom{+}2.8662 \times 10^{-5} & \hspace{15pt} \phantom{+}2.4682 \times 10^{-5} \\
\phantom{+}3.7411 \times 10^{-4} & \hspace{15pt} -1.3378 \times 10^{-2} & \hspace{15pt} -2.9516 \times 10^{-2} \\
\phantom{+}6.6206 \times 10^{-4} & \hspace{15pt} -3.9877 \times 10^{-2} & \hspace{15pt} -1.1349 \times 10^{-2}
\end{pmatrix} \\
f =& 
\begin{pmatrix}
-5.2086 \times 10^{-4} & \hspace{15pt} \phantom{+}7.4068 \times 10^{-4} & \hspace{15pt} -1.1542 \times 10^{-3} \\
-7.1611 \times 10^{-4} & \hspace{15pt} \phantom{+}4.2458 \times 10^{-5} & \hspace{15pt} \phantom{+}9.8693 \times 10^{-4} \\
-4.3188 \times 10^{-4} & \hspace{15pt} \phantom{+}7.6551 \times 10^{-5} & \hspace{15pt} \phantom{+}3.3605 \times 10^{-5}
\end{pmatrix} \nonumber \\
+ i& 
\begin{pmatrix}
-7.6064 \times 10^{-4} & \hspace{15pt} \phantom{+}5.6229 \times 10^{-4} & \hspace{15pt} -9.0955 \times 10^{-4} \\
-2.0928 \times 10^{-4} & \hspace{15pt} -1.0754 \times 10^{-5} & \hspace{15pt} \phantom{+}3.4779 \times 10^{-4} \\
\phantom{+}7.3913 \times 10^{-4} & \hspace{15pt} \phantom{+}1.5777 \times 10^{-4} & \hspace{15pt} -1.3076 \times 10^{-3}
\end{pmatrix}
\end{align}
\end{itemize}

\subsection{Resonant leptogenesis}
In this section, we provide the Yukawa matrices and bare mass matrix for heavy leptons for the benchmark points given in~\autoref{tab:2}.
\begin{itemize}
\item Benchmark point 4
\begin{align}
M_E = & \begin{pmatrix}
\phantom{+}5.4600 \times 10^{7} & \hspace{20pt} \phantom{+}5.0000 \times 10^{10} & \hspace{30pt} 0 \\
\phantom{+}5.0000 \times 10^{10} & \hspace{20pt} -2.4311 \times 10^{7} & \hspace{30pt} 0 \\
0 & \hspace{20pt} 0 & \hspace{30pt} \phantom{+}1.0000 \times 10^{11}
\end{pmatrix} \\
Y_\ell = & \begin{pmatrix}
\phantom{+}6.3944 \times 10^{-4} & \hspace{15pt} -1.2766 \times 10^{-3} & \hspace{15pt} 0 \\
\phantom{+}7.9636 \times 10^{-3} & \hspace{15pt} -9.1949 \times 10^{-3} & \hspace{15pt} 0 \\
0 & \hspace{15pt} 0 & \hspace{15pt} \phantom{+}1.0102 \times 10^{-1}
\end{pmatrix} \nonumber\\
+ i~ & \begin{pmatrix}
-3.0396 \times 10^{-3} & \hspace{15pt} \phantom{+}1.8326 \times 10^{-3} & \hspace{50pt} 0 \phantom{+++} \\
\phantom{+}6.9558 \times 10^{-3} & \hspace{15pt} -1.0250 \times 10^{-2} & \hspace{50pt} 0 \phantom{+++} \\
0 & \hspace{15pt} 0 & \hspace{50pt} 0 \phantom{+++}
\end{pmatrix} \\
f = & \begin{pmatrix}
-4.1598 \times 10^{-6} & \hspace{15pt} -1.0068 \times 10^{-5} & \hspace{15pt} -6.5894 \times 10^{-6} \\
-4.3928 \times 10^{-6} & \hspace{15pt} \phantom{+}1.9821 \times 10^{-5} & \hspace{15pt} -6.7384 \times 10^{-6} \\
\phantom{+}9.9769 \times 10^{-6} & \hspace{15pt} \phantom{+}6.8231 \times 10^{-6} & \hspace{15pt} -3.7482 \times 10^{-6}
\end{pmatrix} \nonumber\\
+ i~ & \begin{pmatrix}
-7.7060 \times 10^{-6} & \hspace{15pt} -8.1271 \times 10^{-6} & \hspace{15pt} -8.1173 \times 10^{-6} \\
-1.2288 \times 10^{-5} & \hspace{15pt} -4.4691 \times 10^{-6} & \hspace{15pt} \phantom{+}2.6436 \times 10^{-5} \\
\phantom{+}3.4718 \times 10^{-6} & \hspace{15pt} -9.3873 \times 10^{-6} & \hspace{15pt} \phantom{+}3.1070 \times 10^{-5}
\end{pmatrix}.
\end{align}

\item Benchmark point 5
\begin{align}
M_E = & \begin{pmatrix}
-6.1120 \times 10^{7}   & \hspace{15pt} \phantom{+}4.0000 \times 10^{10} & \hspace{15pt} 0 \\
\phantom{+}4.0000 \times 10^{10} & \hspace{15pt} \phantom{+}3.1052 \times 10^{7}  & \hspace{15pt} 0 \\
0                       & \hspace{15pt} 0                        & \hspace{15pt} \phantom{+}1.0000 \times 10^{11}
\end{pmatrix} \\
Y_\ell = & \begin{pmatrix}
-2.5255 \times 10^{-3} & \hspace{15pt} -2.0095 \times 10^{-3} & \hspace{15pt} 0 \\
\phantom{+}1.1503 \times 10^{-2} & \hspace{15pt} \phantom{+}7.8333 \times 10^{-3} & \hspace{15pt} 0 \\
0 & \hspace{15pt} 0 & \hspace{15pt} \phantom{+}1.0102 \times 10^{-1}
\end{pmatrix} \nonumber\\
+ i~ & \begin{pmatrix}
-5.4880 \times 10^{-3} & \hspace{15pt} -4.7387 \times 10^{-3} & \hspace{50pt} 0 \phantom{+++} \\
-1.9057 \times 10^{-4} & \hspace{15pt} \phantom{+}2.1181 \times 10^{-3} & \hspace{50pt} 0 \phantom{+++} \\
0 & \hspace{15pt} 0 & \hspace{50pt} 0 \phantom{+++}
\end{pmatrix} \\
f = & \begin{pmatrix}
-3.4037 \times 10^{-6} & \hspace{15pt} \phantom{+}2.4226 \times 10^{-5} & \hspace{15pt} \phantom{+}1.9598 \times 10^{-5} \\
\phantom{+}1.3533 \times 10^{-5} & \hspace{15pt} -1.3556 \times 10^{-5} & \hspace{15pt} -9.4483 \times 10^{-6} \\
\phantom{+}5.4233 \times 10^{-6} & \hspace{15pt} -2.3759 \times 10^{-5} & \hspace{15pt} -1.8253 \times 10^{-5}
\end{pmatrix} \nonumber\\
+ i~ & \begin{pmatrix}
-1.7690 \times 10^{-6} & \hspace{15pt} -3.9296 \times 10^{-6} & \hspace{15pt} \phantom{+}1.6825 \times 10^{-7} \\
-7.3111 \times 10^{-6} & \hspace{15pt} \phantom{+}2.4971 \times 10^{-5} & \hspace{15pt} -2.7061 \times 10^{-6} \\
\phantom{+}2.0088 \times 10^{-5} & \hspace{15pt} -1.5018 \times 10^{-6} & \hspace{15pt} \phantom{+}1.5157 \times 10^{-5}
\end{pmatrix}.
\end{align}

\item Benchmark point 6

\begin{align}
M_E= & \begin{pmatrix}
\phantom{+}1.8023 \times 10^{7} & \hspace{15pt} \phantom{+}3.0000 \times 10^{10} & \hspace{15pt} 0 \\
\phantom{+}3.0000 \times 10^{10} & \hspace{15pt} \phantom{+}1.2090 \times 10^{7} & \hspace{15pt} 0 \\
0 & \hspace{15pt} 0 & \hspace{15pt} \phantom{+}1.0000 \times 10^{11}
\end{pmatrix}, \\
Y_\ell = & \begin{pmatrix}
\phantom{+}2.4951 \times 10^{-4} & \hspace{15pt} \phantom{+}4.8420 \times 10^{-4} & \hspace{15pt} 0 \\
\phantom{+}1.3079 \times 10^{-2} & \hspace{15pt} -6.2182 \times 10^{-3} & \hspace{15pt} 0 \\
0 & \hspace{15pt} 0 & \hspace{15pt} \phantom{+}1.0102 \times 10^{-1}
\end{pmatrix} \nonumber\\
+ i~ & \begin{pmatrix}
\phantom{+}4.1849 \times 10^{-3} & \hspace{15pt} -1.0727 \times 10^{-3} & \hspace{50pt} 0 \phantom{+++} \\
-4.2115 \times 10^{-3} & \hspace{15pt} \phantom{+}1.1493 \times 10^{-3} & \hspace{50pt} 0 \phantom{+++} \\
0 & \hspace{15pt} 0 & \hspace{50pt} 0 \phantom{+++}
\end{pmatrix} \\
f = & \begin{pmatrix}
\phantom{+}4.2527 \times 10^{-6} & \hspace{15pt} -1.6471 \times 10^{-5} & \hspace{15pt} \phantom{+}9.7268 \times 10^{-6} \\
-6.8912 \times 10^{-6} & \hspace{15pt} \phantom{+}2.3502 \times 10^{-5} & \hspace{15pt} \phantom{+}5.7805 \times 10^{-6} \\
-5.2724 \times 10^{-7} & \hspace{15pt} -5.8112 \times 10^{-6} & \hspace{15pt} \phantom{+}9.7668 \times 10^{-6}
\end{pmatrix} \nonumber\\
+ i~ & \begin{pmatrix}
\phantom{+}2.8436 \times 10^{-6} & \hspace{15pt} \phantom{+}4.3370 \times 10^{-6} & \hspace{15pt} -4.2278 \times 10^{-6} \\
\phantom{+}1.0941 \times 10^{-6} & \hspace{15pt} \phantom{+}8.5661 \times 10^{-6} & \hspace{15pt} -4.5486 \times 10^{-5} \\
-6.4069 \times 10^{-6} & \hspace{15pt} \phantom{+}3.4778 \times 10^{-5} & \hspace{15pt} \phantom{+}8.5685 \times 10^{-6}
\end{pmatrix}.
\end{align}

\end{itemize}

%%%%%%%%%%%%%%%%%%%%%%%%%%%%%%%%%%%%%%%%%%%%%%%%%%%%
\end{sloppypar}
\bibliographystyle{utphys}

%\end{sloppypar}
\bibliography{reference}
\end{document}